\documentclass[ ]{elsarticle}
\usepackage{mathtools}
\usepackage{amssymb}

\usepackage{caption}
\makeatletter
\def\ps@pprintTitle{%
 \let\@oddhead\@empty
 \let\@evenhead\@empty
 \def\@oddfoot{}%
 \let\@evenfoot\@oddfoot}

\makeatother
\begin{document}
\journal{ }

\begin{frontmatter}

\title{Metapopulation epidemic models with heterogeneous mixing and travel behaviour}

\author[ad1]{Andrea Apolloni}

\author[ad2,ad3,ad4]{Chiara Poletto}

\author[ad5]{Jos\'e J. Ramasco}

\author[ad6]{Pablo Jensen}

\author[ad3,ad4,ad7]{Vittoria Colizza\corref{cor1}}
\ead{vittoria.colizza@inserm.fr}

\cortext[cor1]{Corresponding author}

\address[ad1]{London School of Hygiene and Tropical Medicine, London, UK}

\address[ad2]{Computational Epidemiology Laboratory, Institute for Scientific Interchange (ISI), Torino, Italy}

\address[ad3]{INSERM, UMR-S 1136, Institut Pierre Louis d'Epid\'emiologie et de Sant\'e Publique,  F-75013, Paris, France}

\address[ad4]{Sorbonne Universit\'es, UPMC Univ Paris 06, UMR-S 1136, Institut Pierre Louis d'Epid\'emiologie et de Sant\'e Publique, F-75013, Paris, France}

\address[ad5]{Instituto de F\'{\i}sica Interdisciplinar y Sistemas Complejos IFISC (CSIC-UIB), Campus UIB, Palma de Mallorca, Spain}

\address[ad6]{Institut des Syst\'emes Complexes Rh\^{o}ne-Alpes (IXXI) and Laboratoire de Physique, UMR 5672, \'Ecole Normale Sup\'erieure de Lyon, 69007 Lyon, France}

\address[ad7]{Institute for Scientific Interchange (ISI), Torino, Italy}

\begin{abstract}
Determining the pandemic potential of an emerging infectious disease and how it depends on the various epidemic and population aspects is critical for the preparation of an adequate response aimed at its control. The complex interplay between population movements in space and non-homogeneous mixing patterns have so far hindered the fundamental understanding of the conditions for spatial invasion through a general theoretical framework. To address this issue, we present an analytical modelling approach taking into account such interplay under general conditions of mobility and interactions, in the simplifying assumption of two population classes.

We describe a spatially structured population with non-homogeneous mixing and travel behaviour through a multi-host stochastic epidemic metapopulation model. Different population partitions, mixing patterns and mobility structures are considered, along with a specific application for the  study of the role of age partition in the early spread of the 2009 H1N1 pandemic influenza.

 We provide a complete mathematical formulation of the model and derive a semi-analytical expression of the threshold condition for global invasion of an emerging infectious disease in the metapopulation system. A  rich solution space is found that depends on the  social partition of the population, the pattern of contacts across groups and their relative social activity, the travel attitude of each class, and the topological and traffic features of the mobility network. Reducing the  activity of the less social group and reducing the cross-group mixing are predicted to be the most efficient strategies for controlling the pandemic potential in the case the less active group constitutes the majority of travellers. If instead traveling is dominated by the more social class, our model predicts the existence of an optimal across-groups mixing that maximises the pandemic potential of the disease, whereas the impact of variations in the activity of each group is less important. 

The proposed modelling approach introduces a theoretical framework for the study of infectious diseases spread in a population with two layers of heterogeneity relevant for the local transmission and the spatial propagation of the disease. It can be used for pandemic preparedness studies to identify adequate interventions and quantitatively estimate the corresponding required effort, as well as in an emerging epidemic situation to assess the pandemic potential of the pathogen from population and early outbreak data. 
\end{abstract}

\begin{keyword}
metapopulation models \sep
\sep epidemic spreading 
\sep complex networks 
\sep mobility 
\sep mixing patterns
\sep travel behaviour 

\end{keyword}

\end{frontmatter}

\section{Background}

The spatial spread of directly transmitted infectious diseases depends on the interplay between local interactions among hosts, along which transmission can occur, and  dissemination opportunities presented by the movements of hosts among different communities. The availability of increasingly large and detailed datasets describing contacts, mixing patterns, distribution in space and  mobility of hosts have enabled a quantitative understanding of these two factors~\cite{chowell03,barrat04,brockmann06,eagle06,gonzalez08,mossong08,balcan09,liljeros01,salathe10,isella11,roth12} and led to the development of data-driven mechanistic models  to capture the epidemic dynamics of  infectious diseases~\cite{eubank04,ciofidegliatti08,halloran08, balcan09}.

Although numerical simulations have crucially contributed to our current ability to explain observed spatial epidemic patterns, predict future epidemic outcomes and evaluate strategies for their control, analytical methods offer an alternative valuable avenue for the assessment of an epidemic scenario that is able to clearly identify the key mechanisms at play and 
shed light on some of the complexity inherent in data-driven approaches. In the context of models for spatially transmitted infectious diseases, the metapopulation approach offers a theoretical framework that explicitly maps the spatial distribution of host population and mobility~\cite{hanski04,may84, grenfell97,keeling02}, while offering a tractable system under certain approximations~\cite{colizza07b,colizza08}.  Originally introduced in the field of ecology and evolution~\cite{hanski04}, it considers a population subdivided into discrete local communities, where the infection transmission dynamics is described through standard compartmental schemes, coupled by connections representing the movements of hosts. Despite the mathematical complexity of explicitly considering the spatial dimension and non-trivial topologies connecting local communities, epidemic metapopulation approaches have shown their ability to analytically explain the failure of feasible mobility restriction measures~\cite{colizza07b,colizza08,bajardi11}, alert on the possible negative impact that adaptive travel behaviour of individuals may have on epidemic control~\cite{meloni11}, and  interpret pathogen competition in space~\cite{poletto13a}.

Based on network theory and reaction-diffusion approaches, these studies have quantified the  potential for a global epidemic to occur in terms of a mathematical indicator, $R_*$~\cite{colizza07b,colizza08}, measuring the average number of subpopulations that an infected subpopulation may transmit the disease to, through mobility of infectious individuals during the outbreak duration. Values larger than 1 indicate that transmission can spatially propagate in the metapopulation system and reach global dimension, whereas epidemics with $R_*<1$ are contained at the source. Different mobility modes, traffic dynamics and path choices have been explored so far within the metapopulation framework~\cite{colizza07b,colizza08,balcan11,belik11,meloni11,liu13,poletto12}, however all these properties have been considered at aggregated fluxes level, implicitly assuming that all individuals resident in the same location are indistinguishable and equivalent. Therefore individuals are also considered homogeneous in their mixing pattern. 

Empirical studies of social and contact networks relevant for disease transmission have however identified several heterogeneities in specific features at the individual or group level~--~including, e.g., the number of contacts, their frequency and duration, contacts' clustering,  assortativity, and their structure into communities~--~that affect the dynamics and control of infectious diseases~\cite{liljeros01,lloyd01,pastor01,schneeberger04,keeling05,lloyd-smith05, meyers05,mossong08,read08,smieszek09, rohani10,salathe10,stehle11,karsai11,rocha11}. A particularly efficient theoretical framework that takes into account variations in population features is the transmission matrix approach that divides the population into groups and considers inter-group heterogeneities~\cite{anderson92, diekmann90,brauer08}. Individuals within the same group are assumed to be homogeneous with respect to their ability to contract and transmit the disease, and this approach can be used when variations at the individual level are considered to be negligible within the same group. Its advantage is to allow for a full parameterization of the model with  the information available from empirical studies and for a mathematical formulation for the  analytical computation of important epidemic parameters and observables, such as the basic reproductive number (measuring the average number of secondary cases per primary case)~\cite{diekmann90}, the final size of the epidemic~\cite{brauer08} and its extinction probability~\cite{nishiura11}.

Although interactions between individuals of different types and at different scales through mobility have been included in numerical approaches, and each of them has been separately addressed in mathematical approaches, their joint integration into a general theoretical framework has yet to be developed. A clear example of the importance of both aspects acting together on the dynamics of an epidemic spreading through a population was recently put forward by the 2009 H1N1 pandemic outbreak, where age was observed to be a relevant factor differentiating between local community outbreaks (mainly driven by children) and case importation into unaffected regions (mainly driven by adults)~\cite{nishiura10a,lam2011,apolloni13}. Broken down to the basic mechanisms at play, the observed pattern could be explained through the interplay between two classes of individuals~--~children and adults~--~having different mixing behaviours~\cite{wallinga06,mossong08} and travel habits~\cite{apolloni13}. Other classifications of the population may be also relevant for the spatial spread of an infectious disease and the risk of an epidemic invasion, as prompted by the empirically observed dependence of travel frequency and contact patterns on different features of the population~\cite{isella11,tripstats}. 

In the present study, we present a general theoretical framework for the assessment of the pandemic risk for an infectious disease spreading through a spatially structured population characterized by contact and mobility heterogeneities.  We integrate  the metapopulation framework with the transmission matrix approach using a parsimonious model based on the subdivision of the population into two groups for each local community. We consider different types of mixing patterns across classes to provide a fundamental analytical understanding of the dependence of the global invasion parameter $R_*$ on epidemiological parameters and population features. By restricting to two classes, it is possible to provide a complete mathematical formulation of the model and recover an equation for $R_*$ that can be solved numerically, with approximate analytical solutions being possible under limit conditions on the parameters. These theoretical results are further tested against  mechanistic Monte Carlo simulations of the infection dynamics in the metapopulation system individually tracking hosts in time and space. The framework is completely general and can be applied to different social settings, where host partition may depend on demographic or socio-economic factors, or to roles/conditions of individuals in specific settings (e.g. health-care workers and patients in hospitals~\cite{isella11}, students classified by gender or class and teachers in schools).

\section{Model description} \label{sec:model}

\begin{figure}[!ht]
\centering
\includegraphics[width=0.7\textwidth,keepaspectratio]{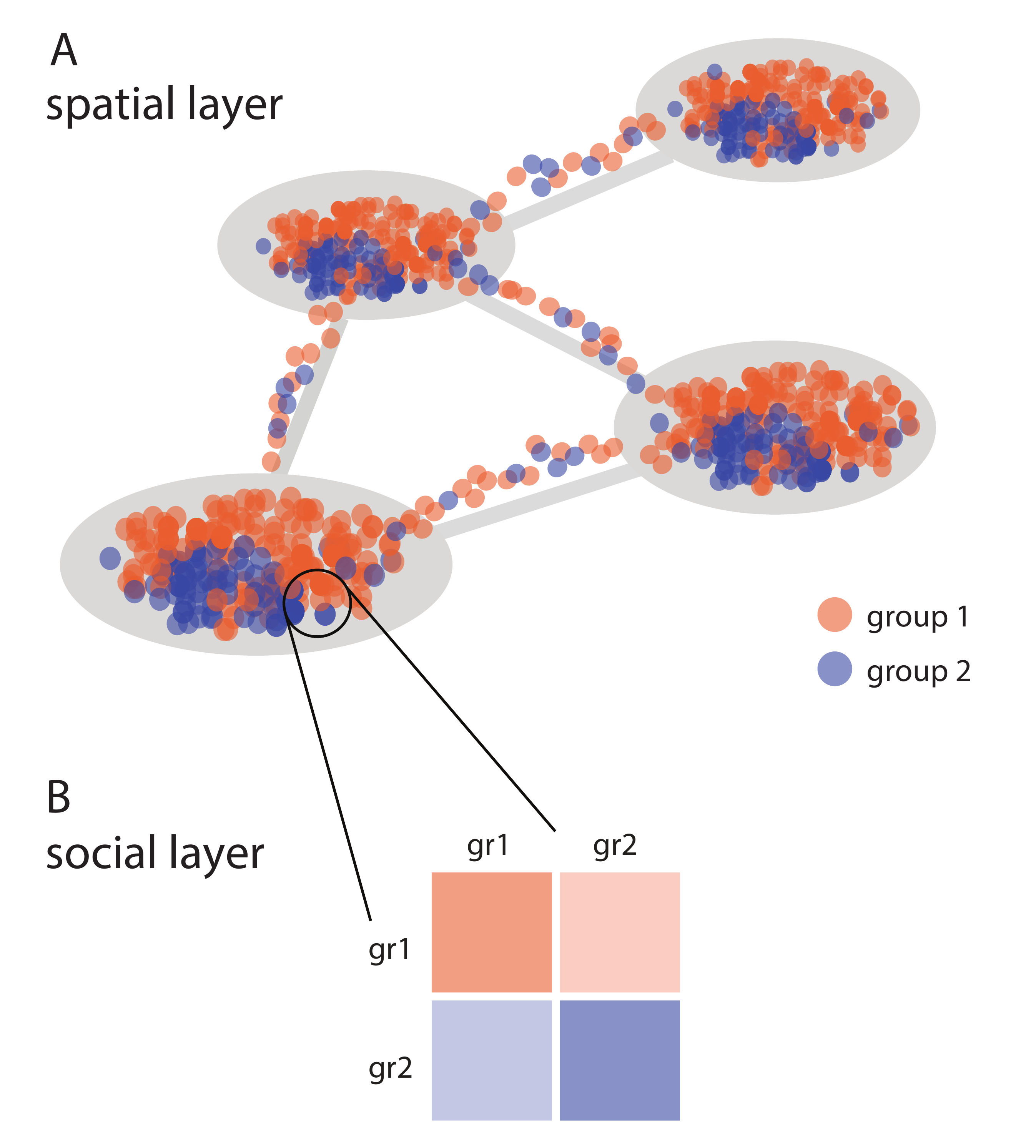}
\caption{\textbf{Scheme of the model.} (A) The spatial layer, based on the metapopulation approach, describes the space structure and the mobility of individuals. (B) The social layer describes  the contact structure within each subpopulation.\\}
\label{fig:model_scheme}
\end{figure}

The modelling approach is based on a metapopulation scheme  where individuals are distributed in subpopulations, or patches, connected by a network of mobility flows (Figure~\ref{fig:model_scheme}).

 It can be described as the integration of two distinct layers: a \emph{social layer}, accounting for heterogeneities in the contact structure among individuals and a \emph{spatial layer}, modelling the distribution of individuals in space and their mobility. Epidemic dynamics occurs inside each patch and is ruled by a transmission matrix approach accounting for the different contact properties of the social classes considered. Mobility properties per class are accounted for in the modelling of individuals movements from one patch to another. In the following we present the two layers in detail, along with the models for  infectious disease transmission and for mobility.

\subsection{Social layer and infectious disease transmission model}

We consider a population socially stratified in two types of individuals (groups), $1$ and $2$,  differing in contact and travel behaviour. We indicate with $\alpha$ the proportion of individuals of type $1$ ($0< \alpha<1$), so that group sizes are given by $N_{l,1}=\alpha\, N_l$ and $N_{l,2} = (1-\alpha)\, N_l$, where $N_l$ is the total number of individuals in a given subpopulation $l$. Interactions among groups can be described by a $2 \times 2 $ contact matrix  encoding the average behaviour of the two groups (in the following we drop the $l$ suffix of the subpopulation under study to simplify our notation) \cite{anderson92} : 
\begin{equation}
\pmb{C}=\left (
\begin{array}{cc}
C_{11} & C_{12}\\
C_{21} & C_{22}\\
\end{array}
\right)= 
\left (
\begin{array}{cc}
\frac{p_1 \, q_1}{\alpha }& \frac{(1-p_2)\, q_ 2}{\alpha }\\
\frac{(1-p_1)\, q_1}{1-\alpha} & \frac{p_2 \, q_2}{1-\alpha }
\end{array}
\right),
\label{contactmatrix2}
\end{equation}
where $C_{ij}$ stands for the contact rate of individuals of type $i$ with those of type $j$  that can be expressed in terms of $q_i$, representing the average number of contacts  per unit time established by an individual of type $i$, and $p_i$,  representing the fraction of those contacts occurring with individuals of the same type. $q_i$ measures the overall social activity of the  group $i$, whereas $p_i$ quantifies how this social activity is distributed among the two groups. Asymmetry in the social activity can be expressed in terms of a parameter $\eta$:
\begin{equation}
\eta=\frac{q_2}{q_1}\,.
\nonumber
\end{equation} 
Interactions are reciprocal in that the number of contacts between individuals of group $1$ and individuals of group $2$ is the same as the number of contacts between group $2$ individuals and group $1$ individuals, requiring the matrix to be symmetric, i.e. $C_{ij} = C_{ji}$. This corresponds to the following condition to be satisfied:
\begin{equation}
(1-\alpha) \,(1-p_2) \, \eta = \alpha \, (1-p_1) \,  \equiv \epsilon, 
\label{relpapb}
\end{equation}
where the parameter  $\epsilon$ here defined quantifies the degree of mixing in the way links are established across classes. It is defined in the range $0 < \epsilon < \text{min} \{ \alpha, \eta \,(1-\alpha)\}$, where values of $\epsilon$ close to zero indicate assortativity of the system (i.e. a tendency of individuals in a given class to preferably interact with individuals of the same class), whereas the upper bound of the range describes a scenario where individuals tend to avoid making contacts within their group. Far from the extremes we have a random or \emph{proportionate} mixing where individuals distribute randomly their contacts in the population. 

The matrix of Eq.~(\ref{contactmatrix2}) can be rewritten as a function of $\eta$, $\alpha$ and $\epsilon$ as:
\begin{equation}
\pmb{C}=q_1 \,\left (
\begin{array}{cc}
\frac{\alpha-\epsilon}{\alpha^2}& \frac{\epsilon}{\alpha \, (1-\alpha)}\\
\frac{\epsilon}{\alpha \, (1-\alpha)}&  \frac{\eta(1-\alpha)-\epsilon}{(1-\alpha)^2}
\end{array}
\right).
\label{contactmatrixPoly}
\end{equation}
Without loss of generality, we consider that individuals in the group $1$ are on average more social than those in group $2$, so that the parameter $\eta$ is defined within the $[0,1]$ interval.
This simplified theoretical framework can be calibrated to describe a real social system, once empirical data on demography and contact behaviour among given classes are available. An example in which individuals are stratified by age is discussed in the Section (\ref{sec:results_h1n1}). A list of all variables used to define the population classes is reported in Table~\ref{summary1}. 

\begin{table}[h!]
\centering

\begin{tabular}{lll}
\\
\hline
\bf{Variable} &\bf{Definition}  & \bf{Range} \\
\hline
$\alpha$ & \shortstack[l]{group 1 fraction of the \\ population} & ]0;1[  \\
\hline 
$q_1,q_2$ & \shortstack[l]{ average number of contacts \\ established by individuals \\ in group 1 and 2}  \\
\hline 
$\eta=\frac{q_2}{q1}$ & \shortstack[l]{ratio of the average number \\ of contacts} & ]0;1] \\
\hline 
$\epsilon$ & \shortstack[l]{total fraction  of contacts \\ across groups} & ]0;min($\alpha,\eta(1-\alpha$))]  \\
\hline
$r$ & \shortstack[l]{group 1 fraction of traveling \\ population} & ]0;1] \\
\hline
\end{tabular}
\caption{Population groups variables \newline}
\label{summary1}
\end{table}

Disease transmission is modelled with a Susceptible-Infectious-Recovered (SIR) compartmental scheme~\cite{anderson92}. Susceptible individuals may contract the infection from infectious individuals and enter the infectious compartment; all infectious individuals then recover permanently and enter the recovered compartment. We indicate with $\beta$ and $\mu$ the transmission rate and the recovery rate, respectively. The infection dynamics is  described by the next generation matrix {\bf \emph{R}}$=\{ R_{ij}\}$ \cite{diekmann90} 
representing the average number of secondary infections of type $i$ generated by primary case of type $j$ in a completely susceptible population. If we assume that disease transmission may only occur along the contacts described by the matrix {\bf \emph{C}}$=\{ C_{ij}\}$, then we can express the next generation matrix as a function of the $C_{ij}$ entries:

\begin{equation}
\begin{split}
\pmb{R} & =  \frac{\beta}{\mu} \pmb{\Gamma} \cdot \pmb{C}= \frac{\beta}{\mu} 
\left(
\begin{array}{cc}
C_{11}\alpha & C_{12}\alpha \\
C_{21} (1-\alpha)   & C_{22}(1-\alpha)\\
\end{array}
\right)\\
\\
& =  \frac{\beta \, q_1}{\mu}
\left(
\begin{array}{cc}
1-\frac{\epsilon}{\alpha} & \frac{\epsilon}{1-\alpha}\\
\frac{\epsilon}{\alpha} & \eta -\frac{\epsilon}{1-\alpha}\\
\end{array}
\right)
\end{split}
\label{NextGenPoly}
\end{equation}
where the matrix $\pmb{\Gamma}$, is a diagonal matrix whose entries correspond to the relative sizes of the groups.
The basic reproductive number $R_0$ is calculated as the largest eigenvalue of the matrix  {\bf \emph{R}}~\cite{diekmann90} and it provides 
a threshold condition for a local outbreak in the community; if $R_0 > 1$ the epidemic will occur and will affect 
a finite fraction of the local population, otherwise the disease will die out.

If we consider an epidemic with $R_0>1$, the final fraction $z_i$ of infected individuals in each group (also called epidemic size) can be calculated for the two types of individuals ($i=1,2$) as the solution of the following coupled transcendental equations~\cite{ball1993}:
\begin{equation}
1-z_i= e^{-\sum_{j} R_{ij} \, \, z_j }. 
\label{eq:finalsize}
\end{equation}

\subsection{Spatial layer and mobility model}\label{sec:spatial_layer}

The spatial component of the model is based on the metapopulation approach. Individuals are divided into $V$ subpopulations, called also patches, or nodes of the mobility network. We assume that all subpopulations of the system are characterised by the same social and demographic features in terms of the two groups introduced, so that the parameters $\alpha$, $\eta$ and $\epsilon$ are  homogeneous across the system. This assumption allows us to treat the problem analytically, however it can be easily relaxed in the numerical simulations. Population size and connectivity of the patches are instead heterogenous quantities. Each subpopulation $l$ has  $N_l$ inhabitants and $k_l$ connections through mobility to other subpopulations (also called degree of the node). The mobility network  is characterised by a random connectivity pattern described by an arbitrary degree distribution $P(k)$. In the following we will explore the role of realistic heterogeneous network structures, adopting power-law degree distributions $P (k) \propto k^{-\gamma}$ that was found to well reproduce the behaviour of human mobility patterns at different spatial levels~\cite{chowell03, barrat04, brockmann06, gonzalez08,balcan09}. Traffic along the links is also heterogeneously distributed. In particular the average number of people $w_{l m}$ travelling along a link from a subpopulation $l$ to a subpopulation $m$ is defined according to the following scaling property observed in real-world mobility data~\cite{barrat04}:
\begin{equation}
w_{lm}=w_0 (k_l k_m')^\theta\,,
\end{equation} 
where $k_l$ and $k_m$ represent the degrees of the two ending nodes, and $\theta$ is system-dependent ($\theta\simeq0.5$ in the worldwide air transportation network~\cite{barrat04}). Travellers are chosen randomly in the origin subpopulation,  the traveling rate being simply defined as $d_{l m}  = w_{l m}/N_l$, however we need to take into account that the two social groups have different attitudes towards mobility. We thus introduce a parameter $r$ indicating the fraction of  individuals of type 1 among the $w_{lm}$ travellers, and express the traveling rates of the two groups as:\begin{equation}
\begin{split}
d_{lm,1} & = r \frac{w_0 (k_l\, k_m)^\theta}{N_{l,1}} = \frac{r}{\alpha} \; d_{lm} \,,\\
d_{lm,2} & = (1-r) \frac{w_0 (k_l\, k_m)^\theta}{N_{l,2}} = \frac{1-r}{1-\alpha} \; d_{lm} \,.
\end{split}
\label{d12}
\end{equation}
The full list of variables used to define the metapopulation model is provided in Table~\ref{summary2}.

\begin{table}[h!]
\centering
\begin{tabular}{lll}
\\
\hline
\bf{Variable} &\bf{Definition}   & \bf{\shortstack[l]{Value used \\in numerical \\simulations}} \\
\hline
$k$ & \shortstack[l]{degree of a subpopulation,\\i.e. number of connections \\ to other subpopulations }& [1;$\sqrt{V}]$   \\
\hline
$P(k)=k^{-\gamma}; \gamma$ & \shortstack[l]{subpopulation  degree \\ distribution; power-law \\exponent }& $\gamma=2.3, 3$   \\
\hline
$V;V_k$ & \shortstack[l]{ total number of \\subpopulations; number \\of subpopulations  \\with degree $k$} &  $V= 10^4$\\
\hline
\shortstack[l]{ $\bar{N},N_{k}=\frac{\bar{N}k^\phi}{\langle k^\phi\rangle}$; \\ $\phi$;\\ $w_0$} & \shortstack[l]{ average population \\of a node, \\ population of a node \\ with degree $k$;\\ power-law exponent; \\mobility scale} & \shortstack[l]{$\bar{N}=10^4$\\ $\phi=3/4$\\ $w_0=0.05$}\\
\hline
\shortstack[l]{$w_{lm}=w_0 (k_l k_m)^\theta$; \\ $\theta$} &\shortstack[l]{ number of travelers  from \\ a subpopulation \\with degree $k_l$\\ to a subpopulation \\ with degree $k_m$; \\power-law exponent} &  $\theta$=0.5\\
\hline
\end{tabular}
\caption{Metapopulation model variables \newline}
\label{summary2}
\end{table}

\section{Analytical treatment and results} \label{sec:results_th}

Identifying and understanding the conditions for the spatial invasion of an infectious disease, once it emerges in a given population or community of individuals, requires the consideration of all scales at play in the system. At the local scale, the reproductive number $R_0$ provides a threshold condition for the occurrence of an outbreak locally. At the global scale, however, additional mechanisms need to be considered that may impede the spatial propagation of the disease from the seed of the epidemic to other regions of the system. Even in the case the condition $R_0 > 1$ is satisfied, the epidemic may indeed fail to spread spatially if the mobility rate is not large enough to ensure the travel of infected individuals to other subpopulations before the end of the local outbreak, or if the amount of seeding cases is not large enough to ensure the start of an outbreak in the reached subpopulation counterbalancing local extinction events. It is then possible to identify at the metapopulation scale an additional predictor of the disease dynamics, $R_*$, that defines the condition for spatial (or global) invasion, $R_*>1$~\cite{cross05, ball97, colizza07b, colizza08}, analogously to the reproductive number $R_0$ at the individual level. An analytical expression for $R_*$ has been found in metapopulation models characterized by homogeneous or heterogenous mobility structures and different types of mobility processes: markovian mobility~\cite{colizza07b,colizza08}, adaptive traveling behaviour in response to the pandemic alert \cite{meloni11},  time varying mobility patterns~\cite{liu13},  non-markovian mobility with uniform return rates (i.e. commuting-type of mobility)~\cite{balcan11,belik11}, or with heterogeneous length of stay at destination~\cite{poletto12,poletto13b}. In all cases, the analytical expression of $R_*$ is obtained with a mean-field approximation assuming that all subpopulations with the same degree are statistically equivalent (\emph{degree-block approximation})~\cite{pastor01,colizza07b,colizza08}. This translates in assuming that all features characterising the metapopulation systems (e.g. population size, traveling flux between two subpopulations, in/out traffic of a subpopulation) can be expressed as functions of the degree of the considered subpopulations. While disregarding more specific properties of each subpopulation that may be related for instance to local, geographical or cultural aspects, such assumption is grounded on a large body of empirical evidence obtained from different transportation infrastructures and mobility systems at a variety of scales, pointing to a degree-dependence of average quantities characterising the system~\cite{barrat04,colizza08}. In addition, this simplifying assumption enables an analytical treatment of the problem while accounting for the large degree fluctuations empirically observed in the data~\cite{colizza07b,colizza08}.

Here we consider the same analytical approach adopted in previous works with the aim of exploring the effects of contact and travel heterogeneities in the host population on the invasion potential of an epidemic. We first define the general theoretical framework and present its analytical treatment, and then focus on different cases representing different interaction types between social groups. 

\subsection{General framework} 

Following the approach of~\cite{colizza07b,colizza08}, we describe the disease invasion at the subpopulation level using a branching tree approximation~\cite{ball97}. The invasion process starts from an initial set of infected subpopulations of degree $k$, denoted by $D^0_k$. Before the end of the local outbreak, each of them may infect some of its neighbours, leading to a second generation of infected subpopulations, $D^1_k$. We can generalise the notation by indicating with $D^n_k$ the number of infected subpopulations of degree $k$ at generation $n$. The spatial invasion of the epidemic is then described by the equation relating subsequent generations of infected subpopulations, $D^{n}_k$ and $D^{n-1}_k$:
\begin{equation}
\begin{split}
D^n_k=&\sum_{k'}D^{n-1}_{k'} (k'-1) P(k|k') \prod_{m=0}^{n-1} \left(1-\frac{D^m_k}{V_k}\right)\cdot\\
&\cdot\Omega_{k'k} \left( \lambda_{k'k,1}, \lambda_{k'k,2}  \right).
\label{eq:branching1}
\end{split}
\end{equation}
Here each of the $D^{n-1}_{k}$ subpopulations has $(k'-1)$ possible connections along which the infection can proceed ($- 1$ takes into account the link through which each of those subpopulations received the infection). In order to infect a subpopulation of degree $k$, three conditions need to occur: (i) the connections departing from nodes with degree $k'$ point to subpopulations of degree $k$, as indicated by the conditional probability $P(k|k')$; (ii) the reached subpopulations are not yet infected, as indicated by the probability $1-D^{n-1}_k/V_k$; (iii) the outbreak will be seeded in the new population with probability $\Omega_{k'k} \left( \lambda_{k'k,1}, \lambda_{k'k,2}  \right)$. The latter term is the one that relates the dynamics of the local infection at the individual level to the coarse-grained view that describes the disease invasion at the metapopulation level. It accounts for the contribution of the two classes of individuals, thus including the effects of non-homogeneous travel behaviours and mixing patterns. The number of infectious individuals of each class moving from a subpopulation with degree $k'$ to a subpopulation with degree $k$ during the entire duration of the outbreak is given by:
 \begin{equation}
\begin{split}
 \lambda_{kk',1} & =d_{kk',1} \frac{z_1 \, N_{k',1}}{\mu} = r \, d_{kk'} \frac{z_1 \, N_{k'}}{\mu} \\
 \lambda_{kk',2} &= d_{kk',2} \frac{z_2 \, N_{k',2}}{\mu} = (1-r) \, d_{kk'} \frac{z_2 \, N_{k'}}{\mu} ,
   \label{lambdas2}
   \end{split}
 \end{equation} 
where $z_1$ and $z_2$ are the epidemic sizes in a single population, as computed by Eq.~(\ref{eq:finalsize}), and $\mu^{-1}$ is the average time during which an individual is infectious, hence the individual can seed the disease in a new population in case of travel. We indicate with $\pi_1$ ($\pi_2$) the extinction probability associated to $\lambda_{kk',1}$ ($\lambda_{kk',2}$) infected individuals seeding  a fully susceptible population. Assuming that the seeding processes of the two classes are independent, the outbreak probability $\Omega_{k'k} \left( \lambda_{k'k,1}, \lambda_{k'k,2}  \right)$ is given by 
\begin{equation}
\Omega_{k'k} \left( \lambda_{k'k,1}, \lambda_{k'k,2}  \right)=1-\pi_1^{\lambda_{k'k,1}}\, \pi_2^{\lambda_{k'k,2}} .
\label{surviving}
\end{equation}
The extinction probabilities are determined by the contact patterns of each type of individuals within the subpopulation. Under the assumption that the infectious period is the same for all hosts, $\pi_1$ and $\pi_2$ can be obtained by solving the following quadratic equation  \cite{adke64,griffiths73,nishiura11}:
\begin{equation}
\pi_i=\frac{1}{1+R_{1i} (1-\pi_1)+R_{2i} (1-\pi_2)},
\label{eq:extinctionprob}
\end{equation}
where the index $i$ refers to the two types of individuals ($i=1,2$) and $R_{ij}$ are the terms of the next generation matrix of Eq.~(\ref{NextGenPoly}). If the infection is not able to produce an outbreak in a single population ($R_0<1$), the only solution is $\pi_1=\pi_2=1$, that is, the epidemic dies out. Otherwise, Eq.~(\ref{eq:extinctionprob}) have solutions in the domain of values $(0,1)$ for $\pi_1$ and $\pi_2$, yielding a non zero probability of global outbreaks. Notice that in the case the system is socially homogenous and there is only one type of individuals the two probabilities reduce to $1/R_0$. 

Eq.~(\ref{eq:branching1}) can be simplified  under the following assumptions: (i) the mobility network is uncorrelated, namely $P(k'|k)= k'P(k')/\langle k \rangle$~\cite{barrat08}; (ii) few subpopulations only are infected, i.e. $D^{n-1}_k / V_k \ll 1$, a good approximation of the state of the system during the initial phase of the outbreak; and (iii) the system is very close to the local epidemic threshold,  i.e. $R_0 - 1 \ll 1$. We first notice that the third assumption implies $\pi_{1,2} \simeq 1$ that allows the linear expansion of  Eq.~(\ref{surviving}) into the following expression:
\begin{equation}
 \begin{split}
& \Omega_{k'k} \left( \lambda_{k'k,1}, \lambda_{k'k,2}  \right)  \simeq (1-\pi_1)\, \lambda_{kk',1}+(1-\pi_2) \,\lambda_{kk',2} = \\
&=\left [ (1-\pi_1) \, r \, z_1 + (1-\pi_2)\, (1-r) \, z_2 \right ]\, \frac{w_0}{\mu} \, (k \, k')^\theta.
\label{surviving_linearly}
 \end{split}
 \end{equation}
 By plugging Eq.~(\ref{surviving_linearly}) into the Eq.~(\ref{eq:branching1}) we obtain:
\begin{equation}
\begin{split}
 D_k^n = & \left[(1-\pi_1) \, r \, z_1 +(1-\pi_2)\, (1-r)\, z_2\right]\, \\
 &\frac{w_0}{\mu}\, \frac{k P(k)}{\langle k \rangle}\, \sum_{k'} D_{k'}^{n-1} \,(k'-1) \, (k k')^\theta .
 \label{eq:branching2}
\end{split}
\end{equation}
By multiplying both sides of the above equation by $k^\theta(k-1)$ and summing over all values of $k$, we obtain a recursive equation
in terms of the functional term $\Theta^n=\sum_k k^\theta (k-1)D^n_k$~\cite{colizza07b,colizza08}:
\begin{equation}
\Theta^n=R_* \, \Theta^{n-1} ,
\end{equation}
where $R_*$ encodes the global invasion threshold for the epidemic to occur. The condition $R_*>1$ guarantees indeed the growth of the number of infected subpopulations in the system and therefore the spatial spread of the epidemic. From Eq.~(\ref{eq:branching2}) we derive the explicit form for $R_*$:  
\begin{equation}
R_* = \left[(1-\pi_1) \, r\, z_1+(1-\pi_2)\, (1-r)\, z_2\right]\,\frac{w_0}{\mu} \chi,
\label{Invasion2}
\end{equation}
where $\chi$ is a combination of moments of the degree distribution of the system  encoding the information on  mobility fluxes and topology:
\begin{equation}
\chi= \frac{\langle k^{2+2 \theta}\rangle-\langle k^{1+2 \theta}\rangle}{\langle k \rangle}.
\end{equation} 
If we assume that the parameters characterising social interactions and travel behaviour are uniform across all subpopulations, the social and spatial layers of the system factorize.  $R_*$ can be then evaluated by computing the combination of moments $\chi$, and solving numerically  Eq.~(\ref{eq:finalsize}) and Eq.~(\ref{eq:extinctionprob}) for the epidemic sizes $z_{1,2}$ and the probabilities $\pi_{1,2}$ respectively. Differently from previous works focusing on homogeneous populations of hosts, an explicit analytical solution of $R_*$ cannot be recovered in the general case, due to the $z_{1,2}$ and $\pi_{1,2}$ terms, however special cases can be solved through series expansion as discussed in the following subsections. 

The global invasion parameter $R_*$ quantifies the potential for the spreading at the spatial level of a specific infectious disease in a given social, demographic and mobility setting and  it can thus be used to provide an estimate of the pandemic risk associated to an emerging epidemic. As an example, we address in Section \ref{sec:results_h1n1}  the case of the 2009  H1N1 influenza pandemic in Europe, highlighting the important role of age classes in determining local transmission and spatial spread of the disease.

Here we focus on a generic partition of the population into two groups and explore the impact of the various ingredients of the system (social, demographic, mobility, and disease ingredients) on the global invasion threshold $R_*$. 
\begin{figure}[!ht]
\centering
\includegraphics[width=0.6\textwidth,keepaspectratio]{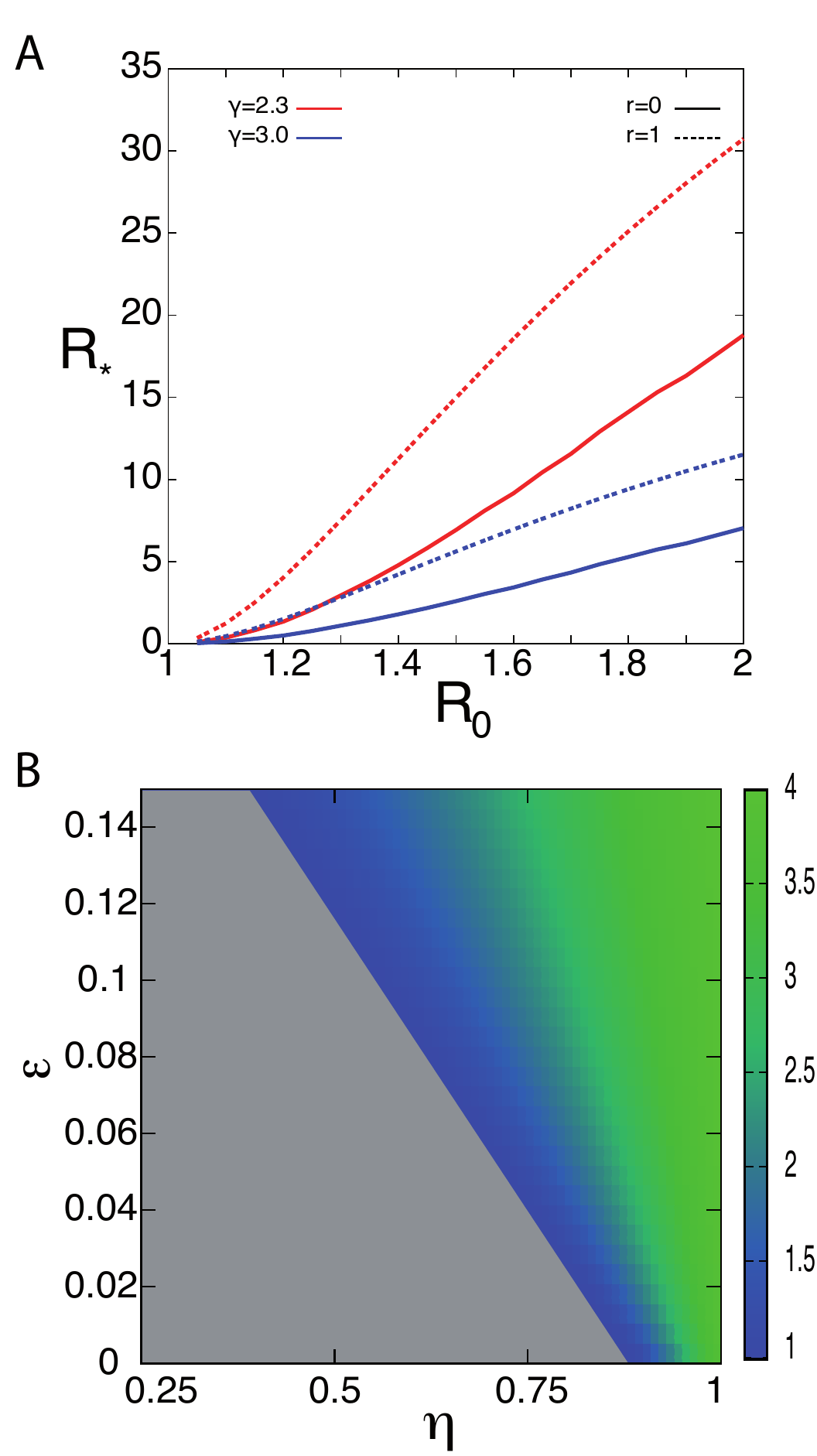}
\caption{\textbf{Numerically computed invasion threshold parameter $R_*$.} (A) $R_*$ as a function of $R_0$ for two different values of the parameter $\gamma$ ruling mobility network heterogeneity ($\gamma=2.3$ and $\gamma=3$) and for boundary values of the traveling partition, $r=0$ and $r=1$. Here we consider a recovery rate $\mu=0.5$, a traffic rescaling factor $w_0=0.05$, and  parameters $\alpha$, $\eta$ and $\epsilon$  set  to 0.2, 0.5, 0.1, respectively. (B) Heat map of $R_*$ as a function of $\epsilon$ and $\eta$ for $\alpha= 0.4$, $R_0= 1.2$ and $\gamma=2.3$. We consider $r=0$. The colour code is proportional to the value of $R_*$, the region of no-invasion $R_*<1$ being  coloured in grey.}
\label{fig:R*}
\end{figure}

Figure~\ref{fig:R*}A shows the dependence of $R_*$ on the reproductive number $R_0$ for different levels of heterogeneity of the human mobility networks, as indicated by the parameter $\gamma$, and considering  two boundary scenarios, $r=0$ and $r=1,$ corresponding to the cases in which only individuals of one group (group 2 or 1, respectively ) travel.  $R_*$ is an increasing function of $R_0$ and assumes larger values for larger heterogeneities in the mobility network (i.e. smaller values of $\gamma$), confirming the results obtained on socially homogenous systems \cite{colizza07b,colizza08}. Moreover, $R_*$ assumes values roughly 50\% larger in the case $r=1$ with respect to the case $r=0$, highlighting the role of different travel behaviour in a partitioned population. When $r$ assumes its boundary values only one group is allowed to travel, whereas the other does not move from the origin subpopulation. If $r=1$, this corresponds to let the most socially active group to travel, thus increasing the probability to start an outbreak at the reached subpopulation, and overall increasing the pandemic potential of the disease considered. This simple result highlights the importance of the characterisation of the passengers profile, in that it may strongly affect the probability of global invasion. 

The role of local contact structure is investigated in Figure~\ref{fig:R*}B. Given a reproductive number $R_0>1$ ensuring the occurrence of a local outbreak in the seeding region, our results show that there exist a region of values of the parameters $\eta$ and $\epsilon$ for which containment at the source is predicted (grey area). Low enough values of the social activity of group 2 vs. group 1 (measured by $\eta$) coupled with large enough assortativity (i.e. low enough values of $\epsilon$) do not provide the conditions for the spatial invasion of the disease.

A more extensive characterisation of the global invasion threshold can be obtained for two specific social systems for which approximate analytical expression of Eq.~(\ref{Invasion2}) can be obtained. We discuss these systems in the following subsections.

\subsection{Proportionate mixing} \label{sec:prop_mixing}

We indicate with \emph{proportionate mixing} the case in which individuals are heterogenous in terms of social activity, but distribute their contacts among the two groups in an unbiased way. As such this model represents the simplest framework to be adopted for describing social stratification \cite{brauer08}, in the case heterogeneities on social activity of individuals are documented but no information on the distribution of  across-group contacts is available \cite{mossong08}. The number of social encounters an individual of group $i$ makes with an individuals of group $j$  is simply determined by the proportion of social contacts of group $j$ with respect to the total number of contacts made by the whole population. Since the number of contacts made by group $i$ per unit time is $q_i \, N_i$, proportionate mixing imposes an extra condition on the probability $p_i$ of internal contacts:
\begin{equation}
p_i= \frac{q_i \, N_i}{q_1\, N_1+ q_2 \,N_2} .
\end{equation}
This condition must be fulfilled together with the symmetry relation of Eq. (\ref{relpapb}). Both conditions translate, in turn, into a relation between the parameters $p_1$, $p_2$, $\alpha$ and $\eta$:
\begin{align}
p_1&=\alpha \, \mathcal D , \nonumber\\
p_2&=\eta \,(1-\alpha)\,\mathcal D,
\label{eq:p_prop}
\end{align}
where $\mathcal D= \left(\alpha+(1-\alpha)\eta\right)^{-1}$. By referring to expression of the contact matrix of Eq.~(\ref{contactmatrixPoly}), the two relations  written in Eq.~(\ref{eq:p_prop}) yield a condition for $\epsilon$, which is not in this case a free parameter but is given by:
\begin{equation}
\epsilon= \eta \,\alpha \,(1-\alpha) \,\mathcal D\,.
\label{eq:eps_prop}
\end{equation}
Notice that, being $\epsilon$ constrained by Eq.~(\ref{eq:eps_prop}), the other two parameters $\alpha$ and $\eta$ can now take values freely in the range $[0,1]$ without any inconsistency in the model.  The contact matrix can be rewritten as:
\begin{equation}
\pmb{C}=
\frac{q_1}{N} \mathcal D
\left (
\begin{array}{cc}
1& \eta\\
\eta & \eta^2\\
\end{array}
\right)\,.
\label{contactmatrixprop}
\end{equation}
From {\bf \emph{C}}, we then derive the next generation matrix: 
\begin{equation}
R=  \frac{ \beta}{\mu} \, q_1 \, \mathcal D \left(\begin{array}{cc}
\alpha& \alpha\, \eta\\
(1-\alpha)\, \eta & (1-\alpha)\, \eta^2\\
\end{array}
\right) .
\label{NextGenprop}
\end{equation}

The calculation of  the epidemic size becomes easier for the proportionate mixing case, as the  relation  $z_2=1-(1-z_1)^\eta$ is satisfied~\cite{brauer08}. Close to the epidemic threshold, where $R_0 \simeq 1$ and $z_{1,2}$  are vanishing, we can write  $z_2\approx \eta \, z_1 +\eta\, (1-\eta)\, z_1^2/2$ and obtain the following expression from Eq. (\ref{eq:finalsize}): 
\begin{equation}
z_1 \approx \frac{2 \,\left(R_0-1 \right) \, \left(\alpha+(1-\alpha)\, \eta^2\right)}{R_0\, \left( R_0 \, (\alpha+(1-\alpha) \, \eta^2)-(1-\alpha)\, (1-\eta)\, \eta^2 \right)}\,.
\label{eq:solhom}
\end{equation} 

The expressions for $\pi_i$ cannot be obtained in a close form. Still, a series expansion provides an approximate solution for the cases $\eta \to 0$ and $\eta \to 1$. The details of the calculations are reported in the \ref{app:prop_matrix}. The first case, $\eta \to 0$, corresponds to a population partition in which the less active group, group 2 in our framework, is fairly isolated and establishes very few contact links. The invasion threshold parameter can be expressed in this case as: 
\begin{equation}
\begin{split}
R_* \simeq &  \, \frac{2\left(R_0-1 \right)^2}{R_0^2} \, \frac{w_0}{\mu} \, \chi \,\cdot \\
&  \left[ r + \eta^2 \, -r \eta^2 \, \left( 1- \frac{(1-\alpha)\, (R_0+1)}{\alpha\, R_0}\right)   \right]\,.
\label{eq:prop0}
\end{split}
\end{equation}
In the case $r = 0$, when only individuals of the type $2$ travel, the threshold $R_*$ converges rapidly to zero (the order being $\eta^2$),  implying that the epidemic remains local and no global spread is  possible. On the other hand, if only individuals of type $1$ travel ($r=1$), $R_*$ approaches rapidly $R_*^h=\frac{2\left(R_0-1 \right)^2}{R_0^2} \, \frac{w_0}{\mu} \chi$, that is the expression of the homogenous case where no partition of the population is considered~\cite{colizza08}. This indicates that individuals of group 2 play a negligible role on the spread of the epidemic. 

The case $\eta \to 1$ represents the homogenous limit, as individuals of the two groups have similar contact patterns, therefore the population looses its criterium for partition. Consistently the linear expansion yields the homogeneous solution $R_*^h$ in addition to a linear correction in $(1-\eta)$: 
\begin{equation}
\begin{split}
R_* \simeq &  \, \frac{2\left(R_0-1 \right)^2}{R_0^2} \, \frac{w_0}{\mu} \, \chi \,\cdot \\
 & \left[ 1+(1-\eta)\,\frac{1-2\,\alpha+r-R_0\, (1-r)}{R_0} \right].
 \label{eq:prop1}
 \end{split}
\end{equation}
 
\begin{figure}[!h]
\centering
\includegraphics[width=10cm,height=10cm,keepaspectratio]{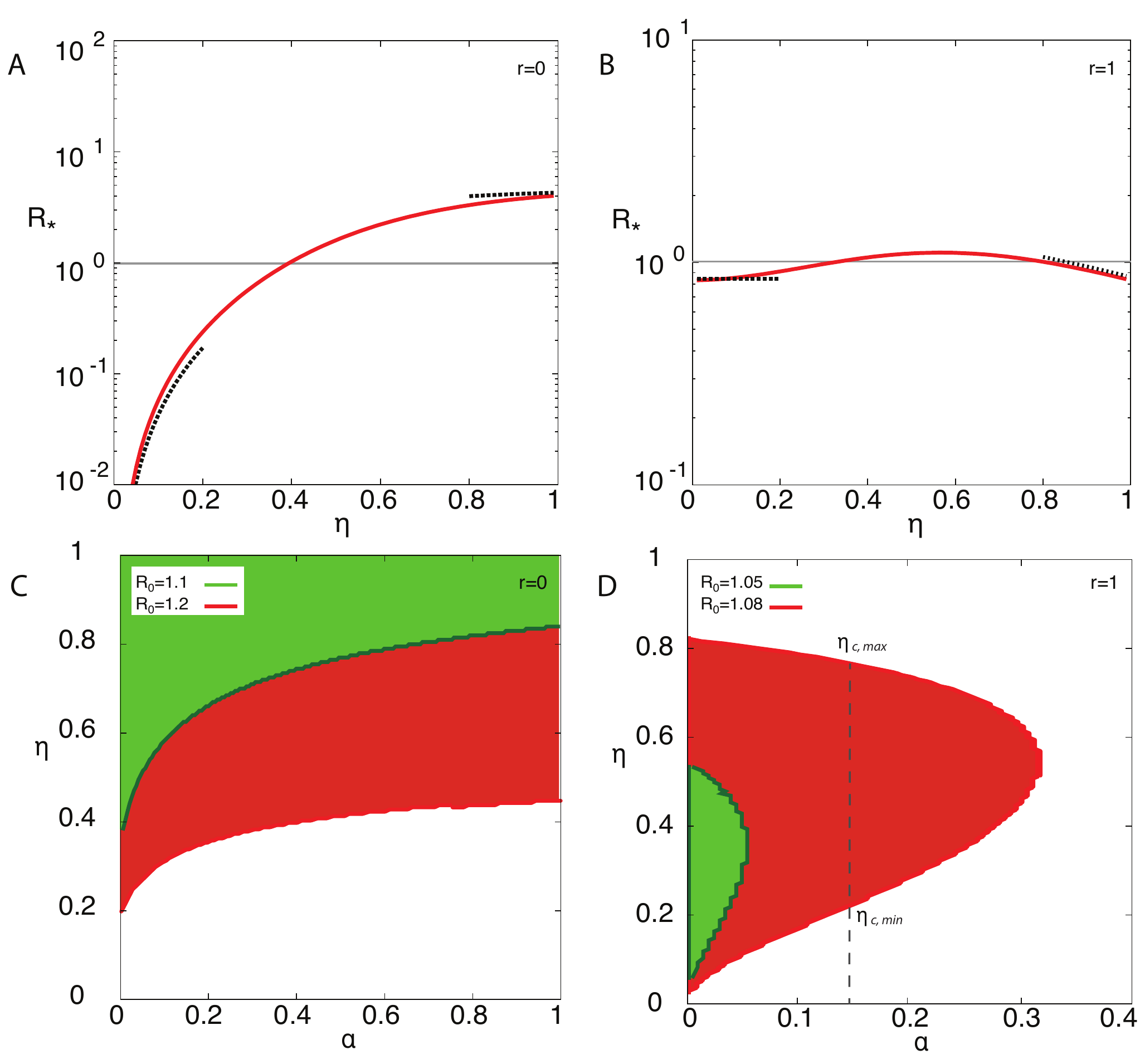}
\caption{\textbf{$R_*$ for a proportionate social system.} On the top $R_*$ as a function of $\eta$. Panel $(A)$ shows the case $r=0$, $\alpha= 0.4$ and $R_0= 1.2$. Panel $(B)$ shows the case $r=1$, $\alpha= 0.4$ and $R_0= 1.08$. The continuous curves represent the value as computed numerically, while the dashed curves represent the approximate solutions for $\eta \to 0$ and $\eta \to 1$.  On the bottom threshold condition $R_*=1$ in the $\alpha$, $\eta$ plane as obtained numerically for different values for $R_0$. Panels $C$ and $D$ consider the cases $r=0$ and $r=1$ respectively. For all the panels $\mu= 0.5$, and the mobility network is characterised by $\gamma= 2.3$ and $w_0=0.05$. The coloured regions are the one for which the invasion condition $R_*>1$ is satisfied. In panel D we also report the $\eta$ range of values $\left[\eta_{c,min}(\alpha),\eta_{c,max}(\alpha) \right]$
for which invasion is obtained for a given value $\alpha$.\\}
\label{fig:R*_prop}
\end{figure}

Figure~\ref{fig:R*_prop} summarises the results of the proportionate mixing case and presents the comparison between the approximate analytical solutions and the numerical ones. Panels A and B show $R_*$ as a function of $\eta$ for the two boundary cases $r=0$ and $r=1$. In the case in which only individuals of group 2 travel ($r=0$),  $R_*$ is very sensitive to variations in $\eta$, spanning several orders of magnitudes when $\eta \in [0,1]$. The parameter $\eta$ characterises the ratio of the social activity of individuals of group 2 (the only seeders in this case) to the one of group 1, thus it determines the contacts that the individuals seeding the infection in a non-infected subpopulation may establish with the population they encounter. Varying its corresponding value strongly affects the probability to observe a global outbreak. On the other hand, when the traveling flux consists only of individuals of group 1, $\eta$ plays a less important role since its variation does not affect the contact pattern of the seeding group, yielding only slight modifications on $R_*$. The approximate analytical solutions of Eqs.~(\ref{eq:prop0}) and (\ref{eq:prop1}) (dashed lines) well reproduce the results obtained numerically.

Panels C and D of Figure~\ref{fig:R*_prop} summarise the impact of the socio-demographic parameters $\alpha$ and $\eta$ on the invasion condition for the two cases  $r=0$ and $r=1$, respectively, and for different values of $R_0$.  The curves represent the invasion threshold condition $R_*(\eta,\alpha)=1$, with the invasion regions located above the curves of panel C, and to the left side of the curves of panel D. If $r=0$, the curve $\eta(\alpha)$ corresponding to the global invasion condition is an increasing function of $\alpha$, indicating that if the fraction of individuals belonging to group 2 is increased, the smaller need to be the associated social activity to reach the outbreak invasion, given that they represent the seeders of the epidemic. If $r=1$, the functional relationship between $\eta$ and $\alpha$ associated with the threshold condition displays a  richer behaviour (panel D). 
In the limits $\eta \to 0$ and $\eta \to 1$, we recover the homogenous mixing regime where, for the two values of $R_0$ considered in the figure, the epidemic is not able to spread globally. If we move from these boundary values to intermediate values of $\eta$, activating the social heterogeneities of the population in the model, we observe an increase in $R_*$ until the invasion threshold is crossed, and global invasion is reached. Differently from the case $r=0$, if $r=1$, i.e. only more active individuals (group 1) travel, the condition $R_*=1$ is not an increasing fraction of $\alpha$. For values of $\alpha$ smaller than a critical value depending on $R_0$, the system experiences invasion for an entire range of $\eta$ values, $\left[\eta_{c,min}(\alpha),\eta_{c,max}(\alpha) \right]$ (panel D). The upper value of this range, $\eta_{c,max}$, becomes larger as the fraction of individuals in group 1 decreases, indicating that even if group 1 is relatively smaller ($\alpha$ decreasing) and less active ($\eta$ increasing), its exclusive dominance on mobility is enough to ensure invasion. Proportionate mixing is then responsible to limit invasion to $\eta \ge \eta_{c,min}(\alpha)$, so that no invasion is obtained by further increasing the social activity of travelers $\eta < \eta_{c,min}(\alpha)$.
 
\subsection{Assortative mixing}

Assortative mixing  represents the case in which individuals interact preferentially within their group, as it applies e.g. to individuals partitioned by age~\cite{mossong08,wallinga06}. Assortativity is mathematically described by the parameter $\epsilon$: when $\epsilon$ is below the value corresponding to the proportionate mixing (Eq.~(\ref{eq:eps_prop})), the system can be said to be assortative. In the following we consider the limit of high assortativity, i.e. the limit $\epsilon \to 0$. We consider moreover the two limits in $\eta$, $\eta \to 0$ and $\eta \to 1$, as before. This allows us to recover the global invasion parameter $R_*$ through series expansion, as detailed in the \ref{app:asso_matrix}. The resulting expressions in the two limits are: 
\begin{equation}
\begin{split}
R_* \simeq & \; \frac{2\left(R_0-1 \right)^2}{R_0^2} \, \frac{w_0}{\mu} \chi \; \Bigg(  r \,+\\ 
& \epsilon^2 \, \frac{\alpha (1-r) \, R_0^2  - (1-\alpha)r \, R_0 + 3 (1-\alpha)r}{\alpha\, (1-\alpha)^2} \Bigg)\,,
\label{Rassor_e0}
\end{split}
\end{equation}
for the limit $\eta \to 0$, and 
\begin{equation}
\begin{split}
R_* \simeq & \; \frac{2\left(R_0-1 \right)^2}{R_0^2} \, \frac{w_0}{\mu} \chi  \; \Bigg( 1- \frac{\epsilon}{\alpha}  \, \frac{R_0-3}{R_0-1} +\\
& (1-\eta)(1-r)\frac{R_0-3}{R_0-1} \Bigg) \,,
\label{Rassor_e1}
\end{split}
\end{equation}
for the limit $\eta \to 1$.

\begin{figure}[!h]
\centering
\includegraphics[width=14.0cm,height=14.0cm,keepaspectratio]{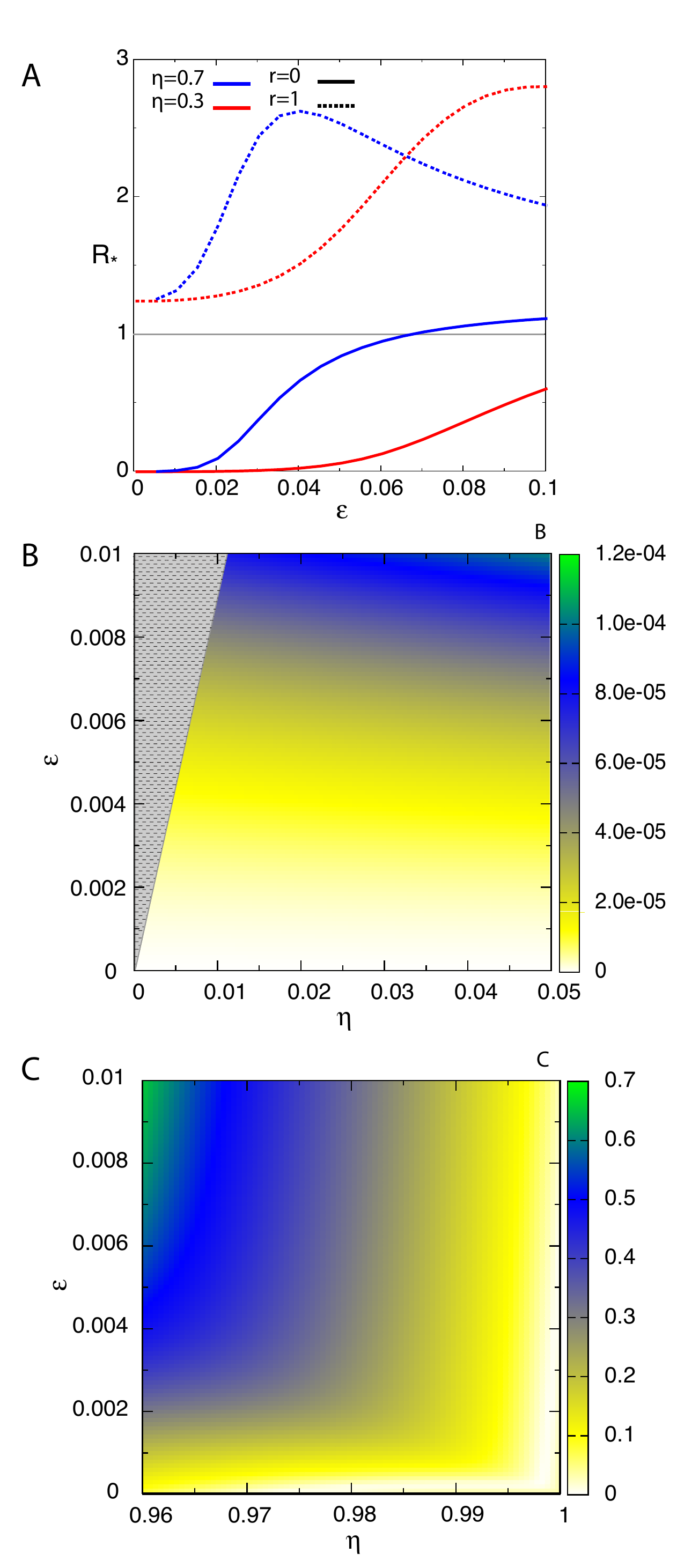}
\caption{\textbf{$R_*$ for an assortative social system.}  (A) $R_*$ as a function of $\epsilon$ for the two cases $r = 0$ and $r = 1$ and two values of $\eta$, 0.3 and 0.7.  $(B)$ Absolute difference between the approximate and the numerically computed value of $R_*$ as a function of $\epsilon$ and $\eta$ for the case $\eta \to 0$. The grey area indicates the parameter region for which the model is not consistent. $(C)$ Absolute difference between the approximate and the numerically computed value of $R_*$ as a function of $\epsilon$ and $\eta$ for the case $\eta \to 1$. In all cases $\alpha=0.1$, $R_0= 1.10$, $\mu= 0.5$, $\gamma= 2.3$ and $w_0=0.05$. \\}
\label{fig:R*_assor}
\end{figure}

Figure~\ref{fig:R*_assor} reports on the results for the assortative mixing case. Panel $A$ shows $R_*$ as a function of $\epsilon$ for the two cases $r=0$ and $r=1$ and for two different values of $\eta$. As for the proportionate mixing case, according to the type of traveling individuals two different behaviours emerge. In the case $r=0$ (continuous curves), $R_*$ is an increasing function of $\epsilon$ and $\eta$. The parameter $\epsilon$ quantifies the chances of cross-group transmission. As such, its increase results in a higher probability for individuals of group 1 to be infected by  imported cases, represented in this case exclusively by individuals of group 2. Being individuals of group 1 more socially active hence more important  for the local spreading, an increase in $\epsilon$ better ensures the occurrence of the outbreak at the local level following importation, and is thus associated to an enhancement in the epidemic invasion potential. On the other hand, when only individuals of group 1 travel ($r=1$, dashed lines in the figure), $R_*(\epsilon)$ is a non monotonous function. Starting from small values of $\epsilon$, the increase in $\epsilon$ favours the global spread (i.e. $R_*$ increases) until a given value is reached, following which a  decrease in $R_*$ is observed. In this case, group 2 only acts in the local transmission dynamics as individuals of the group do not travel ($r=1$). Individuals of group 1 are therefore responsible for the spatial dissemination of the disease and also for the local transmission, being more socially active than the group 2 ($\eta<1$). Our results indicate that there exist an optimal value of the across-groups mixing $\epsilon$ for the assortative case that allows the system to maximise its pandemic potential. A larger number of contacts established between group 1 with respect to the optimal one (i.e. smaller $\epsilon$) would decrease in invasion efficiency because fewer contacts would be directed to the great majority of the population ($\alpha<0.5$), thus reducing the number of infections in the  first group due to interaction with group 2.  An increasingly mixed population (i.e. larger $\epsilon$) would reduce the local spreading role of individuals of class 1 and therefore their capacity to seed other subpopulations. The optimal value of $\epsilon$ clearly depends on all other parameters ($\eta,\,\alpha,\,R_0$).

In panels B and C of Figure~\ref{fig:R*_assor} we show the comparison between the approximate analytical solution and the numerical one by reporting the absolute difference between the corresponding results. The series expansion in Eq.~(\ref{Rassor_e0}) for the limit $\eta \to 0$ yields a quadratic dependence on $\epsilon$ as the first non-constant term, with $\eta$ disappearing from the first two terms of the equation.  The approximated value of $R_*$ so obtained well approaches the numerical results for the case $\eta \to 0$ as shown in panel B where absolute differences are of the order of magnitude of at most $10^{-4}$, and relative differences of at most $\sim 43\%$ in the displayed range. For the limit $\eta \to 1$ we recover instead a linear dependence on the two parameters $\epsilon$ and $\eta$. Panel C of Figure~\ref{fig:R*_assor} shows an absolute difference in $R_*$ below 0.7 between the numerical value and the approximated one,  corresponding to a relative difference of $\sim 36\%$.

\subsection{Proportionate vs. assortative mixing}
\begin{figure}[!h]
\centering
\includegraphics[width=0.5\textwidth,keepaspectratio]{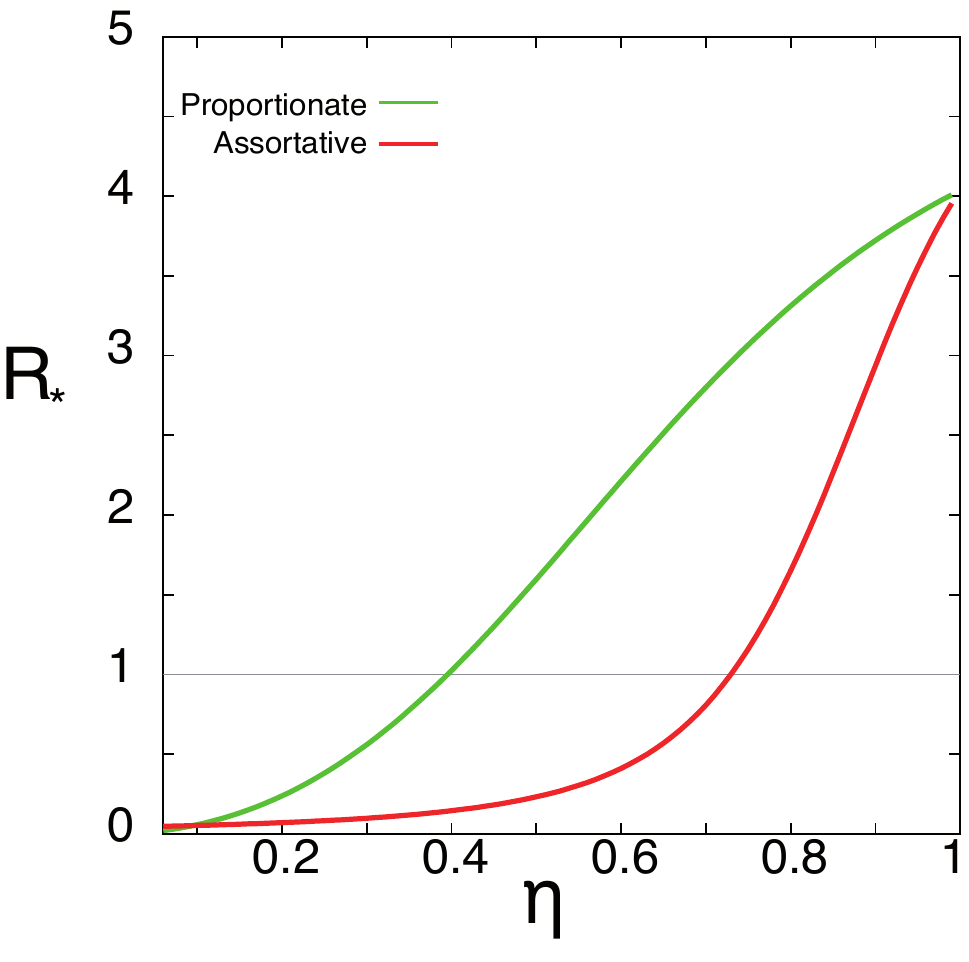}
\caption{\textbf{Comparison between proportionate and assortative social system.} $R_*$ as function of $\eta$ for the proportionate case and the assortative one with $\epsilon= 0.05$. All the other parameters are kept the same in the two curves: $r=0$, $\alpha=0.4$, $R_0= 1.2$,  $\mu= 0.5$, $\gamma= 2.3$ and  $w_0=0.05$.}
\label{fig:R*_prop-ass}
\end{figure}

We conclude this section with a comparison between the proportionate and the  assortative mixing cases. Figure~\ref{fig:R*_prop-ass} shows the value of $R_*$ as a function of $\eta$ for the two cases, proportionate and assortative with degree of across-groups mixing $\epsilon= 0.05$, all the other parameters being equal. Though displaying a qualitatively similar behaviour, the curve obtained in the proportionate mixing case indicates that this specific contact framework favours the global invasion of an emerging infection with respect to the assortative one. Moreover, there exists a range of $\eta$ values for which an epidemic spreading in a population characterized by proportionate mixing would reach a pandemic dimension, whereas the same epidemic would be contained at its source if the population mixes assortatively. Such difference is attributed solely to the different mixing among the two groups.

\section{Numerical simulations} \label{sec:results_sim}

The theoretical framework described so far is based on the combination of continuous differential equations for the transmission dynamics within each subpopulation, with mathematical tools of complex network theory for describing the spatial invasion of the epidemic. In this section we validate the theoretical approach by presenting the comparison between the results recovered so far and the output of stochastic numerical simulations, where all processes are simulated explicitly. The system evolves following a stochastic microscopic dynamics where hosts are individually tracked  and at each time step it is possible to monitor several quantities, as for example the number of infectious individuals within each subpopulation and for each group, or the number of subpopulations reached by the disease. Given the stochastic nature of the dynamics, the experiment can be repeated with different realisations of the noise, different underlying graphs and different initial conditions. 

The mobility network consists of $V=10^4$ subpopulations and is generated by the uncorrelated configuration model~\cite{catanzaro05} that allows building a network with a preassigned degree distribution. In agreement with the analytical calculations we choose a power-law degree distribution, $P(k) \propto k^{-\gamma}$ with exponent $\gamma=2.3$. Once the mobility network is constructed, a number of inhabitants is assigned to each subpopulation according to the degree of the node. Specifically, for each node $l$, we assume a power-law relation between the population $N_l$ and its degree $k_l$, $N_l = \frac{\bar N}{\langle k^\phi \rangle} k_l^\phi$, where the $\bar N$ is the average population of the nodes, set to $10^4$, and $\langle k^\phi \rangle=\sum_k k^\phi P(k)$. This relation was shown to reproduce the behaviour of empirical systems, with an estimate for $\phi$ of approximately $3/4$~\cite{colizza06}. Fluxes along each mobility link also follow a power-law relation with the degrees of the connected nodes, as described in Section~\emph{Spatial layer and mobility model}, $w_{k_lk_m}=w_0 (k_lk_m)^\theta$ , with $\theta=0.5$ and $w_0=0.05$. With this definition, fluxes are symmetric and do not alter the occupancy number of each subpopulation, thus the system is at  equilibrium with respect to the mobility dynamics. The social layer is constructed by dividing the population of each node into  two groups according to the parameter $\alpha$. The contact parameters $\epsilon$ and $\eta$ define then the contact matrix  ruling the transmission dynamics.

The dynamics proceeds in parallel and considers discrete time steps representing the unitary time scale $t$ of the process. The reaction and diffusion rates are therefore converted into probabilities and at each time step the system is updated by implementing the infection dynamics and the diffusion process. Infection transmission is a binomial process that accounts for the heterogeneity of contacts. The force of infection acting on an individual within the group $i$ in the subpopulation $l$ is calculated by combining the contribution of the infectious individuals belonging to the two groups within the same subpopulation, namely
\begin{equation}
\lambda_i= \frac{\beta}{N_l} \left(C_{i1} I_1 + C_{i2} I_2\right),
\label{eq:force_inf}
\end{equation}
where the transmission rate $\beta$ corresponding to the chosen value for $R_0$ is computed from the largest eigenvalue of the next generation matrix -- see Eq.~(\ref{NextGenPoly}). Recovery from the disease is also a binomial process, with  every infectious individual having at each time step a probability $\mu$ to enter in the recovered compartment. We set $R_0=1.2$ and $\mu=0.5$. The diffusion of individuals is implemented as a multinomial process by accounting  the heterogeneities in individual travel frequency given by Eq.~(\ref{d12}). Throughout this numerical exploration we always assumed that only individuals of group 2 travel, i.e. $r=0$. 

The epidemic is initialised by placing 5 infected individuals per each group within a randomly chosen subpopulation and it is simulated until the extinction of the virus is reached. The fraction of subpopulations reached by the disease $D_\infty/V$ provides a clear quantification of the invasion potential of the disease. We consider the two scenarios introduced in the analytical treatment, the proportionate mixing case and the assortative one, and we provide a comparison between the outcome of the numerical simulations and the corresponding analytical results. 

\begin{figure}[!h]
\centering
\includegraphics[width=0.6\textwidth,keepaspectratio]{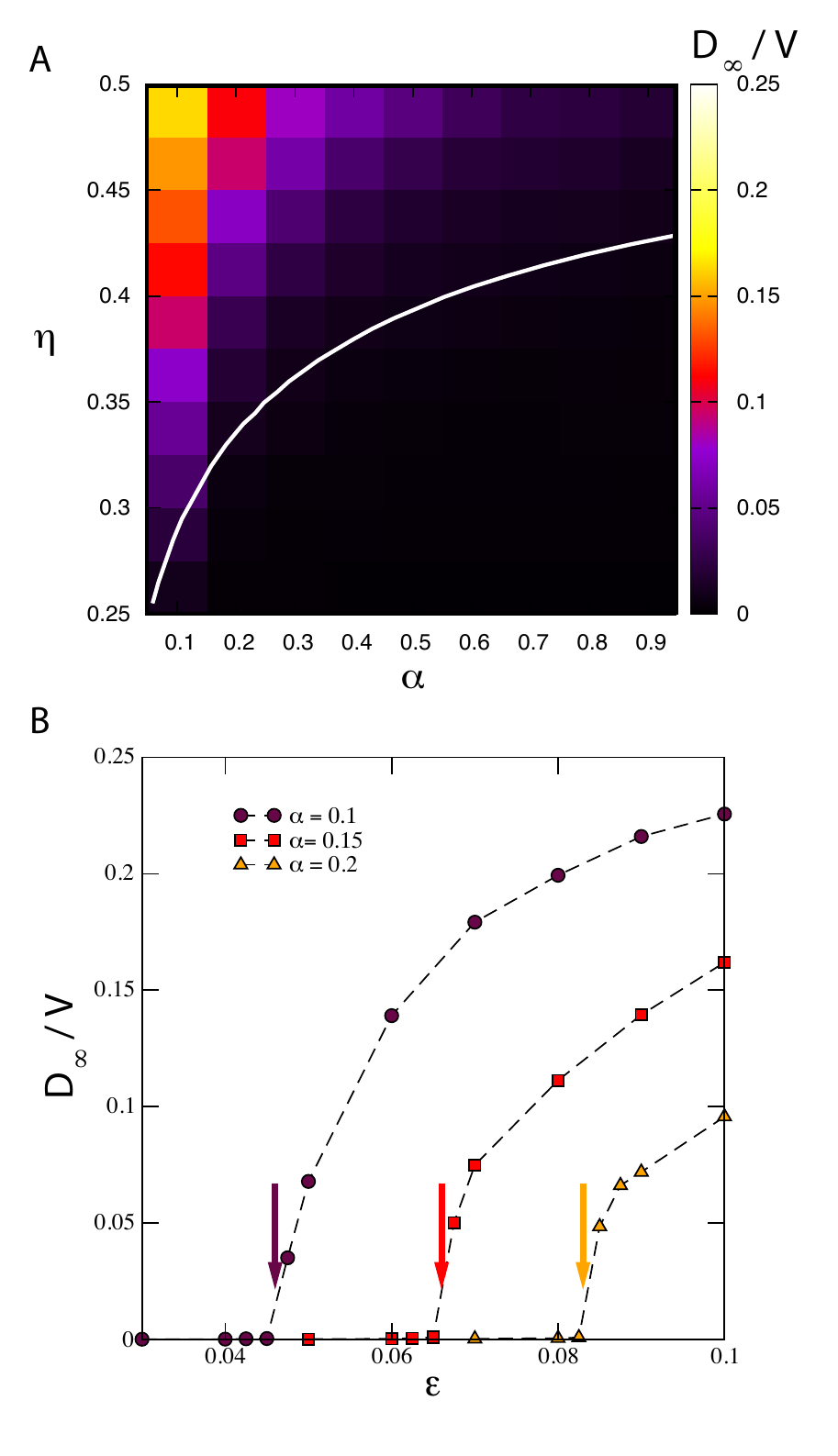}
\caption{\textbf{Comparison between numerical results and analytical estimates.} (A) Invasion behaviour for the proportionate mixing case. $D_\infty/V$ as a function of $\alpha$ and $\eta$ for the case $r=0$. The colour code is proportional to the average value of $D_\infty/V$ as computed from $5000$ stochastic runs. The white line corresponds to the global invasion threshold $R_*(\alpha, \eta)=1$ computed solving numerically the analytical equations. (B) Invasion behaviour for the assortative mixing case. $D_\infty/V$ as a function of $\epsilon$ for $\eta=0.5$ and three different values of $\alpha$, 0.1, 0.15, 0.2. The coloured arrows indicate for the three cases the critical values of $\epsilon$ for which the condition $R_*=1$ is satisfied, as obtained by the analytical equations.} 
\label{fig:sim}
\end{figure}

Panel A of  Figure~\ref{fig:sim} considers the case of proportionate mixing and provides an exploration of the space of parameters $\eta$ and $\alpha$. The heat map shows the average $D_\infty/V$, computed over 5,000 stochastic realisations for each point $(\eta, \alpha)$. The white line indicates the global  invasion threshold  $R_*(\alpha, \eta)=1$ as computed by solving numerically Eq.~(\ref{Invasion2}), in order to allow for a  comparison between the analytical results and the simulations. Notwithstanding finite-size and discrete effects considered in the numerical simulation, and the several approximations used in the analytical treatment (degree-block, branching ratio, and others), the heatmap shows a good agreement between results from simulations and from the numerical solutions of the equations describing the threshold condition for the system. 

Panel B of Figure~\ref{fig:sim} focuses on the assortative mixing case. Here we show the average fraction of infected subpopulations, $D_\infty/V$, as a function of the assortative parameter $\epsilon,$ for three different values of $\alpha$ and for $\eta=0.5$. All the curves present a transition between local outbreak and global invasion in correspondence of a critical value of $\epsilon$, above which  the fraction of infected subpopulation becomes an increasing function of $\epsilon$. The increase in $\alpha$ reduces the invasion potential of the disease. The threshold behaviour is in agreement with the theoretical analysis (Eq.~(\ref{Invasion2})), whose threshold results are reported in the plot for comparison (coloured arrows).

\section{Application to the 2009 H1N1 pandemic influenza} \label{sec:results_h1n1}

The modelling framework introduced so far can provide a prompt scenario analysis in case of an emerging epidemic. Once estimates for the disease parameters are available, the method allows for assessing the invasion potential of the disease for a specific country or region for which data on social contacts and mobility are available. Here we provide as an example the study of the 2009 pandemic of A(H1N1) influenza in Europe and Mexico~\cite{apolloni13}. The relevant partition of the population in this setting is the subdivision in age classes, following the empirical evidence collected during the initial phase of the epidemic. The analysis of early outbreak data indeed showed that the majority of cases due to local transmission in the community was among children, whereas imported cases~--~seeding the epidemic in non-infected areas~--~were mainly adults~\cite{apolloni13,nishiura11,nishiura10a}. Each age class was mainly responsible for one of the two mechanisms at play in the spreading~--~local transmission (children), and spatial dissemination (adults). To explicitly study the role of these two types of hosts on the conditions for global invasion, we consider the generic multi-host metapopulation framework introduced here with an age partition that is parameterized with demographic and contact data. We consider a children age class (group 1) of  individuals below 18 years old and an adult age class (group 2) of the remaining population. The fraction  $\alpha$ of population of group 1 is obtained from UN statistics~\cite{ONU}. The average for Europe is $\alpha= 0.197$ and  other values are reported in Table~\ref{tab:h1n1case}. Contact parameters $\epsilon$ and $\eta$ are estimated  from the contact matrices reconstructed from the large data-collection of the  Polymod project for eight countries in Europe~\cite{mossong08,apolloni13}. The average estimates across the eight countries are $\epsilon= 0.097$ and $\eta=0.795$, and additional estimates for specific countries are reported as examples in Table~\ref{tab:h1n1case}. The European situation is also compared to the one of Mexico~\cite{fraser09_full}, seed country of the pandemic, to explore the impact of very different social contexts on the epidemic dynamics. 

\begin{table}[!h]
\centering
\begin{tabular}{llll}
\hline
\bf{Country} &$\boldsymbol{\alpha}$ & $\boldsymbol{\eta}$ & $\boldsymbol{\epsilon}$ \\
\hline
Germany & 0.183 & 0.746 & 0.098 \\
Netherlands & 0.221 & 0.833 & 0.094 \\
Poland & 0.212 & 0.972 & 0.100 \\
\emph{Europe} & \emph{0.197} & \emph{0.795} & \emph{0.097} \\
\hline 
Mexico & 0.320 & 0.323 & 0.063 \\
\hline
\end{tabular}\caption{Values of parameters $\alpha$, $\eta$ and $\epsilon$ for three European countries~\cite{mossong08}, for the European average~\cite{mossong08,apolloni13}, and for Mexico~\cite{fraser09_full}.\newline}
\label{tab:h1n1case}
\end{table}

The values presented in the table describe an assortative system, where social activity is heterogeneous among the two groups, with children having on average more contacts than adults. Air-transportation statistics available for several airports yield an average of 7\% of children occupancy\cite{apolloni13}, thus  $r=7\%$. Finally we parametrize the mobility network and the distribution of traveling fluxes by setting $\gamma=2.3$ and $w_0=1$~\cite{barrat04}. 

Epidemiological parameters were chosen among the estimates provided for the A(H1N1) pandemic. Throughout the analysis we consider an infectious period of 2.5 days \cite{balcan09} and three different estimates for $R_0$: $R_0=$1.05 (corresponding to the estimate in~\cite{balcan09} for the reproductive number in Europe during summer 2009), $R_0=$1.20 (as estimated from the outbreak data in Japan~\cite{nishiura10b}), and $R_0=1.40$ (as estimated from the early outbreak data in Mexico~\cite{fraser09}). We also consider a scenario in which a certain fraction of the adult population has a pre-existing immunity to the virus accounting in this way for the serological evidence indicating that about 30 to 37\% of the individuals aged $\ge 60$ years had an initial degree of immunity prior to exposure~\cite{ikonen10}. We assume that $33\%$ of individuals aged $\ge 60$ years are immune and completely protected against H1N1 pandemic virus~\cite{apolloni13}, and for each country we compute the corresponding fraction of the adult group with pre-exposure immunity.

\begin{figure}[!h]
\centering
\includegraphics[width=0.5\textwidth,,keepaspectratio]{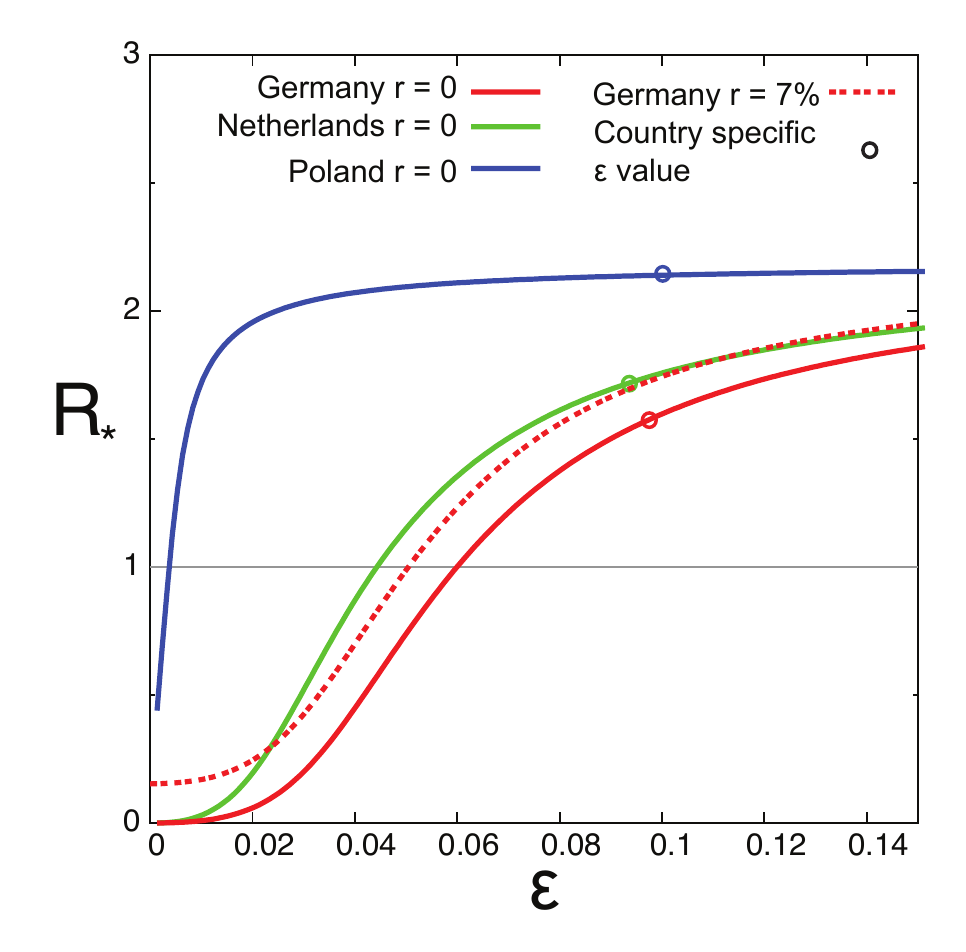}
\caption{\textbf{$R_*$ as a function of $\epsilon$ for the three european countries analysed.} For three cases we set $r=0$. In the case of Germany  we compare the case $r=0$ with $r=0.07$ as estimated by empirical data.} 
\label{fig:R*_h1n1_1}
\end{figure}

With all the parameters being informed by the data, we address the impact of the specific socio-demographic context on the invasion threshold by comparing  three European countries taken as examples (Germany, Netherlands and Poland), along with a comparison Europe vs. Mexico. Figure~\ref{fig:R*_h1n1_1} shows $R_*$ as a function of $\epsilon$ for the three countries assuming $R_0=1.05$. We consider the case $r=0$ for Poland and Netherlands and we compare the two cases $r=0$ and $r=7\%$ for  Germany. The heterogeneities induced by different values of $\alpha$ and $\eta$ may impact significantly the invasion behaviour, as shown by the great discrepancy among the two curves of Germany and Poland: an increase of $\eta$ from 0.75 to 0.97 lowers the critical value of $\epsilon$ for which invasion is reached of more than one order of magnitude. For $\epsilon$ values in this range, the same disease could thus lead to two different scenario (invasion or containment) if emerging in two different countries (Poland or Germany, respectively). Given the values of $\epsilon$ obtained from data of the three countries (Table~\ref{tab:h1n1case}), we obtain that even with very low estimates of the reproductive number, taking into account the seasonal suppression of transmission during summer 2009~\cite{balcan09}, all countries under study are predicted to experience a spatial propagation of the outbreak once seeded, confirming the situation observed in reality.

The comparison between the case $r=0$ and $r=7\%$ for Germany allows us to quantify the role of children as seeders of the epidemic in new locations in a data-driven situation. They contribute to the increase of the invasion potential of the epidemic, thus lowering the minimum value of the across-groups mixing for which the epidemic spatial spread is possible. The effect  is small but appreciable. 

\begin{figure}[!h]
\centering
\includegraphics[width=0.5\textwidth,keepaspectratio]{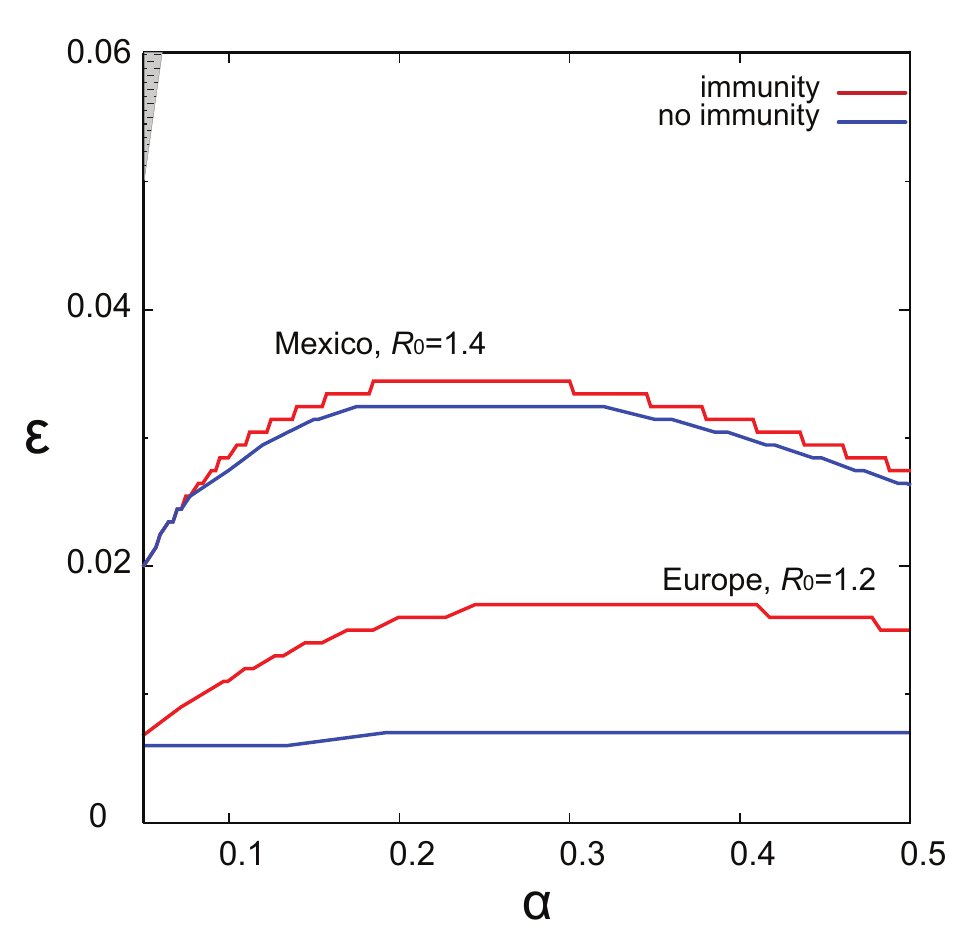}
\caption{\textbf{Threshold condition $R_*=1$  for Europe and Mexico.} Threshold condition $R_*=1$ as a function of $\epsilon$ and $\alpha$ for Europe (bottom curves) and Mexico (top curves): comparison of the no-immunity case with the case of pre-existing immunity. Here we consider: $R_0=1.2$ in Europe and $R_0=1.4$ in Mexico. All travellers are adults ($r=0$). The two lines red and blue correspond to pre-existing immunity and no-immunity. Global epidemic invasion region is above each critical curve. The patterned grey area refers to the region of parameter values that do not satisfy the consistency relation.} 
\label{fig:R*_h1n1_2}
\end{figure}

If we consider pre-existing immunity in the older age classes, we observe how differences in the population demographic profile across different regions of the world may have a strong impact in the resulting suppression of the pandemic potential due to prior immunity. Figure~\ref{fig:R*_h1n1_2} shows the critical curves $R_*=1$  in the $\alpha$, $\epsilon$ plane for Europe and Mexico. As expected, immunity reduces the parameter space leading to global invasion (in each panel, above each critical curve) since a fraction of the population is now modelled to be fully protected against the virus. For a given $\alpha$, a larger mixing across age classes is  needed for the pathogen to spatially propagate in a population having pre-existing immunity; similarly, a more assortative population would be able to contain the disease at the source. It is interesting to note that the magnitude of this effect on the critical curve for invasion is affected by the population profile. The effect is indeed smaller for Mexico than for Europe, since the Mexican population has a smaller percentage of population in the $\ge 60$ class of age with respect to Europe and thus an overall smaller proportion of the population who is fully protected by the pre-existing immunity.

\section{Conclusions}
This study  presented a general  theoretical framework to account for two different layers of heterogeneity relevant for the propagation of epidemics in a spatially structured environment, namely contact structure and heterogenous travel behaviour. The model presents a structure with two distinct scales~--~a social scale and a spatial one. Employing a subdivision into two host classes, we provide a mathematical formulation of the model and derive a semi-analytical solution of the invasion equation, encoding the conditions for the global invasion of the epidemic. The system is characterized by a very rich space of possible solutions, depending on the demographic profile of the population, the pattern of contacts across groups and their relative social activity, the travel attitude of each class, and the topological and traffic features of the mobility network. Two qualitatively different scenarios are found. The increase of the across-group mixing and of the social activity of the less active group (relative to the more active group) enhance the pandemic potential of the infectious disease, if seeders are mostly found in the less active group. Reductions of the number of contacts of individuals of the less active group is predicted to be the most efficient strategy for reducing the pandemic potential. If instead traveling is dominated by the most active class, the role of the contacts ratio between the two groups is negligible for a given population partition, whereas there exist an optimal across-groups mixing that maximizes the pandemic potential of the disease. Reductions or increases of this quantity with respect to the optimal value would decrease the probability that the epidemic, once seeded in a given region, would reach a global dimension. Such findings call for the need to develop further studies to identify appropriate intervention measures that can act on these socio-demographic aspects, depending on the type of partition and of population considered. Empirical data of contact patterns, demography and travel from eight European countries and from Mexico, and of the 2009 H1N1 influenza pandemic were used to parametrize our model in terms of two age classes of individuals~--~children and adults~--~and explain the spatial spread of the disease following emergence (in Mexico) and international seeding (in Europe). Despite the need to address some limitations of the model in future work (e.g. partition in more than two classes, and geographic dependence of population features), our approach offers a flexible theoretical framework~--~validated on historical epidemics~--~that can promptly assess the pandemic potential of an emerging infectious disease epidemic where a specific socio-demographic stratification is relevant in the disease transmission among individuals.

\section*{Acknowledgments}

This work has been partially funded by the ERC Ideas contract no. ERC-2007 -Stg204863 (EPIFOR) and the EC-Health contract no. 278433 (PREDEMICS) to VC and CP; the ANR contract no. ANR-12-MONU-0018 (HARMSFLU) to VC; the Ram\'on y Cajal program and the project MODASS of the Spanish Ministry of Economy (MINECO), and the EC projects EUNOIA and LASAGNE to JJR.



\newpage
\appendix
\section*{Appendices}\label{Appendices}

\section{Series expansion for the proportionate mixing case\label{app:prop_matrix}}

In this section we provide the details of the series expansion in the quantities $z_1$, $z_2$ and $\pi_1$, $\pi_2$ that allow for recovering the approximate results for $R_*$ of Eqs.~(\ref{eq:prop0}) and (\ref{eq:prop1}). 

For the limit $\eta \to 0$, the epidemic size of Eq. (\ref{eq:solhom}) can be approximated by

\begin{align}
z_1	&\simeq 2\, \frac{(R_0-1)}{R_0^2} \left( 1+\eta^2\,  \frac{(1-\alpha)}{\alpha\, R_0 }\right), \nonumber\\
z_2 & \simeq 2\, \eta\, \frac{(R_0-1)}{R_0^2}. 	
\label{z:eta0}
\end{align}

While the extinction probabilities, which are the solutions for Eq. (\ref{eq:extinctionprob}), are

\begin{align}
\pi_1 &\simeq \frac{1}{R_0}\, \left(1 - \eta^2\, \frac{(1-\alpha)\, (R_0-1)}{\alpha}\right), \nonumber \\
\pi_2 &\simeq 1-(R_0-1)\, \eta. 	
\label{pi:eta0}
\end{align}
We combine these expressions into Eq.~(\ref{Invasion2}) and we keep only the first not-constant terms, recovering in this way Eq.~(\ref{eq:prop0}).

For the case $\eta \to 1$ we recover first the solutions for Eq. (\ref{eq:solhom}) in the first order in $(1-\eta)$

\begin{align}
z_1 & \simeq \frac{2\, (R_0-1)}{R_0^2} \, \left( 1+ (1-\eta)\, \frac{(1-\alpha)}{R_0} \right) \nonumber\\
z_2 & \simeq \frac{2\, (R_0-1)}{R_0^2}\, \left( 1+(1-\eta)\, \frac{(1-\alpha -R_0)}{R_0}\right).
\label{z:eta1}
\end{align}
and then the extinction probabilities become

\begin{align}
\pi_1& \simeq\frac{1}{R_0 }\, \left(1-(1-\eta) \, \frac{(1-\alpha)\, (R_0-1)}{R_0}\right)\nonumber \\
\pi_2&\simeq\frac{1}{R_0}\, \left(1+ (1-\eta)\,\frac{\alpha\, (R_0-1)}{R_0}\right). 
\label{pi:eta1}
\end{align}

with these expressions for $\pi_{1,2}$ and together with $z_{1,2}$ of Eq. (\ref{z:eta1}), we find in this limit Eq.~(\ref{eq:prop1}). As expected the first term in Eq.~(\ref{eq:prop1}) is the homogenous solution $R_*^h$, while the second is liner in $1-\eta$. 

\section{Series expansion for the assortative mixing case\label{app:ass_matrix}}

For the case of assortative mixing scenario the contact matrix and next generation matrix are more complex and need to be expanded before handling Eq.~(\ref{eq:finalsize}) and Eq.~(\ref{eq:extinctionprob}). We first consider the series expansion in $\epsilon$ of the basic reproductive number, which up to second order term reads 

\begin{equation}
R_0 \simeq \, \frac{\beta\, q_1}{\mu} \, \left(1+\frac{\epsilon}{\alpha} + \frac{\epsilon^2}{(1-\alpha)\, \alpha\, }\right) .
\end{equation}

We go up to the second order because it is necessary to find the leading terms for $R_*$ in the limit $\eta \to 0$ and $r\to 0$. However, most of calculations are performed up to first order in $\epsilon$ only. The contact matrix can be written in this limit as

\begin{align}
\label{ContactAss}
 & \; \; \; \pmb{C} \simeq \frac{q_1}{N} \times \\ 
 & \left(
\begin{array}{cc}
\frac{1}{\alpha} - \frac{\epsilon^2}{(1-\alpha)\, \alpha^2\, (1-\eta)} & \frac{\epsilon}{\alpha\, (1-\alpha)} \, \left( {\scriptstyle 1+}\frac{\epsilon}{\alpha} \right)\\
\, &\, \\
\frac{\epsilon}{\alpha\, (1-\alpha)} \, \left( {\scriptstyle 1+} \frac{\epsilon}{\alpha} \right)
 & \frac{\eta}{1-\alpha}- \frac{\epsilon \,(\alpha - \eta\, (1-\alpha))}{\alpha\, (1-\alpha)^2}+ \frac{\epsilon^2\,(\eta\, (1-\alpha) -\alpha)}{\alpha^2\, (1-\alpha)^2 \, (1-\eta)} \\
\end{array}
\right) . \nonumber
\end{align}

While the next generation matrix becomes

\begin{align}
\label{NextGenPolyAss}
& \; \; \;  \pmb{R} \simeq R_0 \times \\
& \left(
\begin{array}{cc}
 {\scriptstyle 1 -}\frac{\epsilon^2}{(1-\alpha)\, \alpha\, (1-\eta)} & 
\frac{\epsilon\, (\alpha+\epsilon)}{(1-\alpha)\, \alpha}\\
\, &\, \\
 \frac{\epsilon\,(\alpha+\epsilon) }{\alpha^2} & 
 {\scriptstyle  \eta +\epsilon\,} \left( \frac{\eta}{\alpha} - \frac{1}{1-\alpha}\right) {\scriptstyle   + \epsilon^2\,} \frac{ \eta\, (1-\eta)-\alpha\, (1+\eta-\eta^2)}{(1-\alpha)\, \alpha\, (1-\eta)}  \\
\end{array}
\right) . \nonumber
\end{align}

The solution of the epidemic size for the two groups is obtained by considering first the behaviour of (\ref{eq:finalsize}) around the epidemic threshold, $R_0\approx 1$, and then expanding the solutions up to first or second order with respect to $\epsilon$. In this way, the solutions can be written as a sum of terms in order zero, one and two in $\epsilon$, $z_{1,2} = z_{1,2}^0+\epsilon \, z_{1,2}^1 + \epsilon^2 \, z_{1,2}^2$. As before and for simplicity, we consider two limits regarding the contact rate of individuals of the two types: $\eta \to 1$ and $\eta \to 0$. Note, however, that the limit has to satisfy the condition $\epsilon < \eta \, (1-\alpha)$.

For the case $\eta \to 0$, by taking into account the matrices of Equations (\ref{NextGenPolyAss}) and (\ref{ContactAss}), we can solve Eq.~(\ref{eq:finalsize}) obtaining for the epidemic sizes 

\begin{align}
z_1 \simeq & \,\frac{2\, (R_0-1)}{R_0^2} \nonumber \\ 
&- \epsilon^2 \, \frac{2\,(R_0-1)\, (R_0-2)\,(1+\eta\, (R_0+1))}{\alpha\, R_0^2\,(1-\alpha)}  , \nonumber\\
z_2 \simeq & \, \epsilon \, \frac{2 \,(R_0-1)\,(1+\eta \, R_0) }{R_0\,(1-\alpha)},
\label{finalsizePolye0}
\end{align}

and for the extinction probabilities $\pi_{1,2}$

\begin{align}
\pi_1  \approx & \, \frac{1}{R_0} - \epsilon^2\, \frac{(R_0-1)\, (1+\eta \, (R_0+1))}{\alpha\, R_0\, (1-\alpha)}   ,\nonumber \\
\pi_2  \approx & \, 1- \epsilon\,\frac{(R_0-1)}{1-\alpha} \,(1+\eta\, R_0) .
\label{extprobPolye0}
\end{align}

The terms of the type $2$ individuals appear in Eq.~(\ref{Invasion2}) within the product $(1-\pi_2)\, z_2$. In zero order in $\epsilon$, $\pi_2^0 = 1$ and $z_2^0 = 0$. Therefore, the only terms contributing to $R_*$ in order $\epsilon^2$ of the type 2 individuals are the first order terms $\pi_2^1$ and $z_2^1$. For this reason and for clarity, we have shown in Equations (\ref{finalsizePolye0}) and (\ref{extprobPolye0}) the first order terms for $\pi_2$ and $z_2$ alone. The same cannot be said for the terms of type 1 individuals and so the expansions in $\pi_1$ and $z_1$ are taken up to second order.  Inserting the results of Eqs.~(\ref{finalsizePolye0}) and (\ref{extprobPolye0}) into Eq.~(\ref{Invasion2}), we find the solution of Eq.~(\ref{Rassor_e0}).

In the limit $\eta \to 1$, the solutions for the epidemic sizes in single populations reads

\begin{align}  
z_1  \simeq & \, \frac{2\, (R_0-1)}{R_0^2} - \epsilon \, \frac{2\, (R_0-2)}{\alpha\, R_0^2} ,\nonumber\\
z_2 \simeq & \, \frac{2\, (R_0-1)}{R_0^2} + \frac{2\, (R_0-2)}{R_0^2} \left( (1-\eta) - \frac{\epsilon}{\alpha} \right) .
\end{align} 

The analytical result above points out a peculiar feature of the assortative system this limit condition: small but non-zero outbreaks are possible even for $R_0 = 1$. Such outbreaks have size proportional to a combination of $\epsilon$ and $1-\eta$ which are vanishing quantities \cite{vandendriessche02}. The extinction probabilities in this case are

\begin{align} 
& \pi_1 \simeq \frac{1}{R_0} \, \left( 1 - \frac{\epsilon}{\alpha} \right), \nonumber\\
& \pi_2 \simeq \frac{1}{R_0} \, \left( 1 + (1-\eta) - \frac{\epsilon}{\alpha} \right).
\end{align}

With these expressions for $z_{1,2}$ and $\pi_{1,2}$, we find Eq.~(\ref{Rassor_e1}). In this case we include the linear terms leaving out the sub-leading terms of order $\epsilon^2$ and $\epsilon\, (1-\eta)$.

\newpage
\bibliographystyle{elsarticle-harv}
\bibliography{RD_social}
\end{document}